\documentclass[iop]{emulateapj}   		

\usepackage{subfigure}					
\usepackage{graphicx}					
\usepackage{amssymb}					
\usepackage{mathtools}					
\usepackage{natbib}						

\usepackage{color}						
\usepackage{amsmath}
\usepackage{bm}						
\usepackage{multirow}					

\begin{document}

\title{Using Real and Simulated Measurements of the Thermal Sunyaev-Zel'dovich Effect to Constrain Models of AGN Feedback}

\author{Alexander Spacek$^{1,2}$, Mark Richardson$^{3,4}$, Evan Scannapieco$^1$, Julien Devriendt$^3$, Yohan Dubois$^5$,\\Sebastien Peirani$^{5,6}$, Christophe Pichon$^5$}

\affil{\scriptsize{$^1$School of Earth and Space Exploration, Arizona State University,  PO Box 876004, Tempe, AZ 85287, USA}}
\affil{\scriptsize{$^2$Center for Theoretical Astrophysics, Los Alamos National Laboratory, PO Box 1663, Los Alamos, NM 87545, USA}}
\affil{\scriptsize{$^3$Sub-department of Astrophysics, University of Oxford, Keble Road, Oxford OX1 3RH, UK}}
\affil{\scriptsize{$^4$Astrophysics Department, American Museum of Natural History, Central Park West at 79th Street, New York, NY 10024, USA}}
\affil{\scriptsize{$^5$CNRS and UPMC Universit\'e Paris 06, UMR 7095, Institut d`Astrophysique de Paris, 98 bis Boulevard Arago, Paris 75014, France}}
\affil{\scriptsize{$^6$Universit\'e C\^ote d`Azur, Observatoire de la C\^ote d`Azur, CNRS, Laboratoire Lagrange, France}}


\begin{abstract}

Energetic feedback from active galactic nuclei (AGNs) is often used in simulations to resolve several outstanding issues in galaxy formation, but its impact is still not fully understood. Here we derive new constraints on AGN feedback by comparing observations and simulations of the thermal Sunyaev-Zel'dovich (tSZ) effect. We draw on observational results presented in \citet{Spacek2016, Spacek2017} who used data from the South Pole Telescope (SPT) and Atacama Cosmology Telescope (ACT) to measure the tSZ signal from $\geq 10^{11} M_\odot$ and $\geq 1$ Gyr galaxies at $z$=0.5-1.0 (low-$z$) and $z$=1.0-1.5 (high-$z$). Using the large-scale cosmological hydrodynamical simulations Horizon-AGN and Horizon-NoAGN, which include and omit AGN feedback, we extract simulated tSZ measurements around galaxies equivalent to the observational work. We find that the Horizon-AGN results only differ from the SPT measurements at levels of 0.4$\sigma$ (low-$z$) and 0.6$\sigma$ (high-$z$), but differ from the ACT measurements by 3.4$\sigma$ (low-$z$) and 2.3$\sigma$ (high-$z$). The Horizon-NoAGN results provide a slightly better fit to the SPT measurements by differing by 0.2$\sigma$ (low-$z$) and 0.4$\sigma$ (high-$z$), but a significantly better match to the ACT measurements by differing by only 0.5$\sigma$ (low-$z$) and 1.4$\sigma$ (high-$z$). We conclude that, while the lower-mass ($\lesssim 5 \times 10^{11} M_\odot$) SPT results allow for the presence AGN feedback energy, the higher-mass ($\gtrsim 5 \times 10^{11} M_\odot$) ACT results show significantly less energy than predicted in the simulation including AGN feedback, while more closely matching the simulation without AGN feedback, indicating that AGN feedback may be milder than often predicted in simulations.

\end{abstract}


\section{Introduction}
\label{sec:simintro}

Galaxies are some of the most prominent objects in the Universe, but the processes governing their formation and evolution are surprisingly uncertain. Although early models favored hierarchical galaxy evolution, in which progressively larger galaxies form stars at later times \citep[e.g.][]{Rees1977,White1991}, an increasing amount of observational evidence reveals a more complex history. For example, at later times there is the appearance of cosmic `downsizing' \citep[e.g.][]{Cowie1996}. Since $z\approx2$ the typical mass of star-forming galaxies has decreased by a factor of $\approx$ $10$ or more \citep{Drory2008}. A similar history is observed for AGNs, whose typical luminosities have decreased by as much as a factor of $\approx$ $1000$ since $z\approx2$ \citep{Hopkins2007}. This observed concurrent downsizing trend of both galaxies and AGNs, combined with other well-known relationships between supermassive black holes and their host galaxies \citep[e.g.][]{Shankar2016}, points to a mechanism likely affecting both the small scale of the supermassive black hole ($\lesssim1$ ly) and the large scale of the galaxy ($\gtrsim100$ kly).

One such mechanism is feedback during the AGN phase of supermassive black hole evolution \citep[e.g.][]{Scannapieco2004,Granato2004,Croton2006}. AGNs are energetic enough to drive out enormous, powerful radio jets, as well as extremely luminous radiative winds, causing energetic outflows through the host galaxy. This feedback has the potential to blow out and heat up gas within and around the galaxy, preventing both further star formation in the galaxy and further accretion onto the supermassive black hole. Incorporating AGN feedback into numerical galaxy evolution models has been shown to be very effective in reproducing observed galaxy trends \citep[e.g.][]{Sijacki2007,Lapi2014,Kaviraj2017}, including downsizing \citep[e.g.][]{Thacker2006,Hirschmann2012}. However, the specific details of AGN feedback remain uncertain because precise details are very difficult to measure \citep[e.g.][]{Fabian2012}.

One of the most promising methods for directly measuring the impact of AGN feedback on galaxies and clusters is by looking at anisotropies in the cosmic microwave background (CMB) photons passing through hot, ionized gas. If the gas is moving with a bulk velocity, there will be frequency-independent fluctuations in the CMB temperature known as the kinematic Sunyaev-Zel'dovich (kSZ) effect, while if the gas is sufficiently heated there will be redshift-independent fluctuations in the CMB temperature known as the thermal Sunyaev-Zel'dovich (tSZ) effect \citep[][]{Sunyaev1972}. The tSZ effect can be integrated over a region of the sky to give a direct measurement of the gas thermal energy \citep[e.g.][]{Scannapieco2008}. Measurements of the tSZ effect have been very useful in detecting massive galaxy clusters \citep[e.g.][]{Reichardt2013}. Simulations have also shown that the tSZ effect around galaxies can be effective in distinguishing between models of AGN feedback \citep[e.g.][]{Chatterjee2008, Scannapieco2008}.

Significant observational work has already been done to try and measure the tSZ effect around galaxies and AGNs. For example, \citet{Chatterjee2010} used data from the Wilkinson Microwave Anisotropy Probe and Sloan Digital Sky Survey (SDSS) around both quasars and galaxies to find a tentative $\sim 2\sigma$ tSZ signal suggesting AGN feedback; \citet{Hand2011} used data from SDSS and ACT to find a $\sim 1\sigma-3\sigma$ significant tSZ signal around galaxies; \citet{Gralla2014} used the ACT to find a $\sim 5\sigma$ significant tSZ signal around AGNs; \citet{Greco2015} used SDSS and Planck to find a $\gtrsim 3\sigma$ significant tSZ signal around galaxies; \citet{Ruan2015} used SDSS and Planck to find $\sim 3.5\sigma-5.0\sigma$ significant tSZ signals around both quasars and galaxies; \citet{Crichton2016} used SDSS and ACT to find a $3\sigma-4\sigma$ significant tSZ signal around quasars; and \citet{Hojjati2016} used data from Planck and the Red Cluster Sequence Lensing Survey to find a $\sim 7\sigma$ tSZ signal suggestive of AGN feedback.

Additionally, recent measurements have been made around quiescent, moderate redshift elliptical galaxies \citep[][hereafter tSZ-SPT and tSZ-ACT, respectively]{Spacek2016,Spacek2017}, where a signal due to AGN feedback is expected \citep[e.g.][]{Scannapieco2008}. The signal is very faint, however, so measurements from a large number of galaxies must be stacked. In tSZ-SPT, they performed this type of stacking analysis using the VISTA Hemisphere Survey (VHS) and Blanco Cosmology Survey (BCS) along with the 2011 SPT data release, finding a 3.6$\sigma$ signal hinting at non-gravitational heating based on simple energy models (see Equations (\ref{eq:simEgrav}) and (\ref{eq:simEAGN})). In tSZ-ACT, they used the Sloan Digital Sky Survey (SDSS) and Wide-Field Infrared Survey Explorer (WISE) along with the 2008/2009 ACT data, finding a 1.5$\sigma$ signal consistent with gravitational-only heating based on the same simple energy models.

Directly measuring the energy and distribution of hot gas around galaxies can only reveal so much about the specific physical mechanisms resulting in the observations. In order to place constraints on AGN feedback and other non-gravitational heating processes, it is necessary that observational work be complemented by accurate, detailed simulations. There is a rich history of simulations of the Sunyaev Zel'dovich effect \citep{Scaramella1993,Hobson1996,daSilva2000,Refregier2000,Springel2001,Seljak2001,Zhang2002,Roncarelli2007} and of complementing tSZ measurements and AGN feedback with simulations. For example, both \citet{Scannapieco2008} and \citet{Chatterjee2008} used large-scale cosmological simulations to give predictions for measuring AGN feedback with the tSZ effect; \citet{Cen2015} used simulations to investigate the feedback energies from quasars and their implications for tSZ measurements; \citet{Hojjati2015} used large-scale cosmological simulations to estimate AGN feedback effects on cross-correlation signals between gravitational lensing and tSZ measurements; and \citet{Dolag2016} used large-scale simulations to study the impact of structure formation and evolution with AGN feedback on tSZ measurements.

In this work, we use the large-scale cosmological simulations Horizon-AGN and Horizon-noAGN, which are simulations with and without AGN feedback, respectively \citep{Dubois2014,Dubois2016,Peirani2017}, to complement the work done in tSZ-SPT and tSZ-ACT. We investigate a similar population of moderate redshift, quiescent elliptical galaxies and simulate their tSZ measurements. We then use their measurement distribution and stacking statistics to give insight into the previous observational results. These Horizon simulations have a comoving volume of (100 Mpc/$h$)$^3$, 1024$^3$ dark matter particles, and a minimum cell size of 1 physical kpc. This allows for a large enough population of our galaxies of interest to make robust tSZ measurements, since such high redshift, massive elliptical galaxies are generally uncommon.

The structure of this paper is as follows: in Section \ref{sec:simtsz}, we discuss the tSZ effect and various models of AGN feedback. In Section \ref{sec:simhorizon}, we discuss the Horizon-AGN and Horizon-noAGN simulations. In Section \ref{sec:simdata}, we discuss how we select and measure the tSZ effect around the simulated galaxies. In Section \ref{sec:simmeasure}, we give the parameters and measurements of our selected galaxies. In Section \ref{sec:simdiscussion}, we discuss implications for our results in regard to tSZ-SPT and tSZ-ACT, Horizon-AGN and Horizon-noAGN, and tSZ measurements of AGN feedback in general.

Throughout this work, we use a $\Lambda$ Cold Dark Matter cosmological model with parameters following the Horizon simulations: $h=0.704$, $\Omega_0$ = 0.272, $\Omega_\Lambda$ = 0.728, and $\Omega_b = 0.0455$, where $h$ is the Hubble constant in units of 100 km s$^{-1}$ Mpc$^{-1}$, and $\Omega_0$, $\Omega_\Lambda$, and $\Omega_b$ are the total matter, vacuum, and baryonic densities, respectively, in units of the critical density.


\section{The tSZ Effect}
\label{sec:simtsz}

When CMB photons pass through hot, ionized gas, inverse Compton scattering between the photons and electrons causes the photons to gain energy, and some of these photons are scattered along our line of sight. The resulting shift in the CMB spectrum, known as the tSZ effect \citep[][]{Sunyaev1972}, has a distinctive frequency dependence, with a weaker signal below, and a stronger signal above, 217.6 GHz. This is the frequency where the tSZ effect is null; measurements near 217.6 GHz can be scaled to other frequencies and subtracted to remove contaminating (i.e. non-tSZ effect) signal. The nonrelativistic change in the CMB temperature $\Delta T$ as a function of frequency $\nu$ due to the tSZ effect is
\begin{equation} 
\frac{\Delta T}{T_{\text{CMB}}} = y \left( x \frac{e^x + 1}{e^x - 1} - 4 \right).
\label{eq:simDeltaT}
\end{equation}
Here, $T_{\text{CMB}} = 2.725$ K is the observed CMB temperature today, $x$ is a dimensionless frequency given by $x \equiv \frac{h \nu}{k T_{\text{CMB}}} = \frac{\nu}{56.81 \, \text{GHz}}$, $h$ is the Planck constant, $k$ is the Boltzmann constant, and the dimensionless Compton-$y$ parameter is defined as
\begin{equation}
 y \equiv \int dl \, \sigma_T \frac{n_e k \left( T_e - T_{\rm CMB} \right)}{m_e c^2}, 
 \label{eq:simy}
\end{equation}
 where $\sigma_T$ is the Thomson cross section, $m_e$ is the electron mass, $c$ is the speed of light, $n_e$ is the electron number density, $T_e$ is the electron temperature, and the integral is performed over the line-of-sight distance $l$.
 
\begin{figure*}[t]
\centerline{\includegraphics[height=12cm]{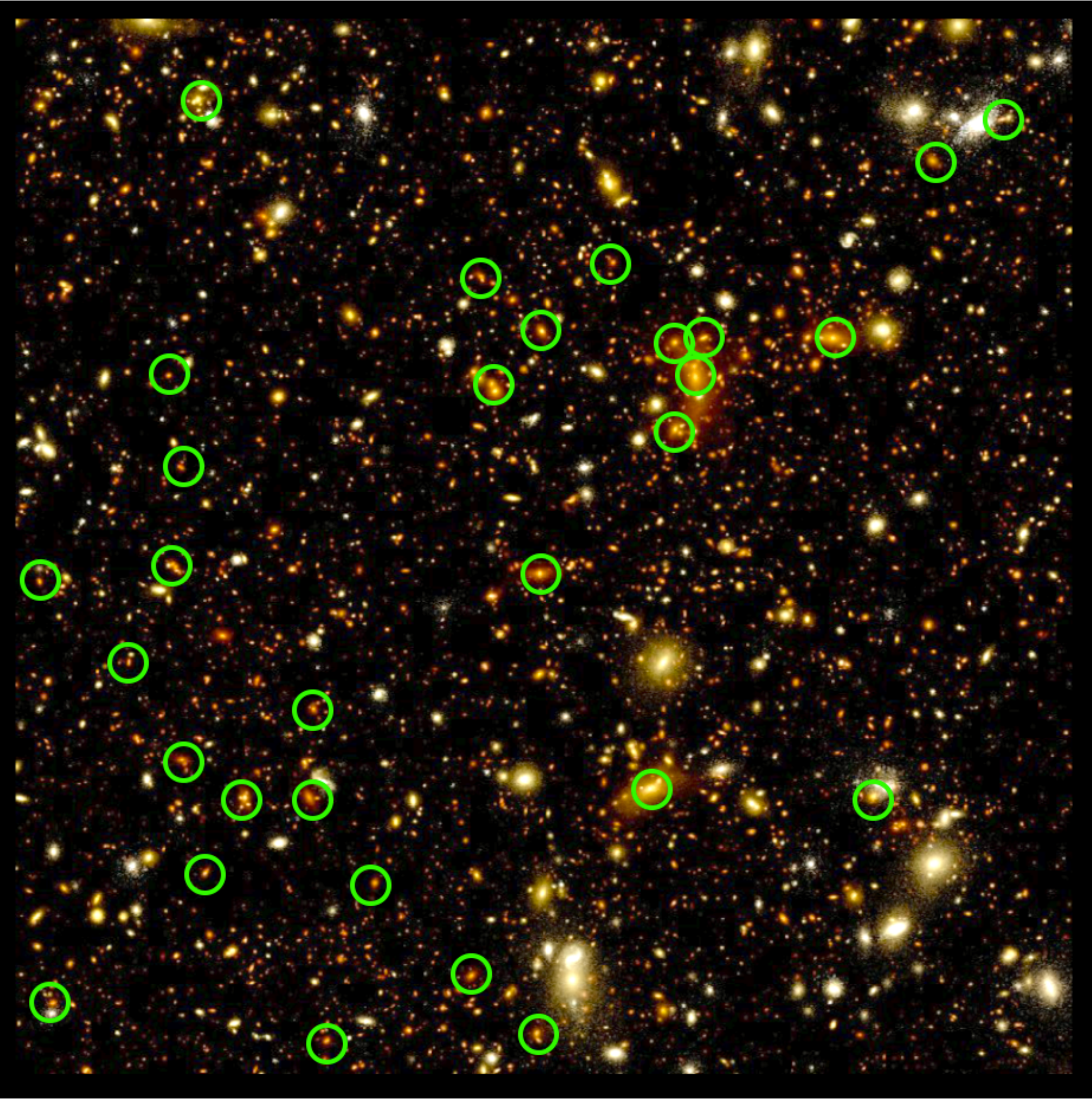}}
\caption{Adapted from \citet{Kaviraj2017}, a 14 arcmin$^2$ simulated $u$, $r$, $z$ image from the Horizon-AGN lightcone \citep{Pichon2010}. The 0.15 arcmin pixel$^{-1}$ resolution image is constructed from star particles within $0.1 < z < 5.8$. Selected galaxies from the sample used in this work are annotated with green circles.}\vspace{2mm}
\label{fig:examplegal}
\end{figure*}
 
\newpage When observing the tSZ effect on the sky, a useful quantity is the angularly integrated Compton-$y$ parameter, $Y$, given by
\begin{equation}
Y \equiv l_{\rm ang}^2 \int y(\bm{\theta}) d\bm{\theta},
\label{eq:simbigy}
\end{equation}
where $l_{\rm ang}$ is the angular diameter distance. For the SPT 150 GHz parameters from tSZ-SPT, this is $Y_\text{SPT} = 2.7 \times 10^{-8}  \, \text{Mpc}^2 E_{\rm 60}$, where $E_{\rm 60}$ is the total line-of-sight gas thermal energy $E_\text{therm}$ in units of $10^{60}$ erg. For the ACT 148 GHz parameters from tSZ-ACT, this is $Y_\text{ACT} = 2.9 \times 10^{-8}  \, \text{Mpc}^2 E_{\rm 60}$. The values are slightly different due to the different frequency sensitivities and beam profiles of the two telescopes. $Y$ can additionally be scaled by redshift and written as
\begin{equation} 
\widetilde{Y} \equiv \frac{Y}{l_{\rm ang}^2} \times E(z)^{-2/3} \times \left(\frac{l_{\rm ang}}{500\ \text{Mpc}}\right)^2 ,
\label{eq:simytilde}
\end{equation}
where $E(z) \equiv \sqrt{\Omega_0 (1+z)^3+\Omega_\Lambda}$ is the Hubble parameter. It often appears this way in the literature \citep[e.g.][]{PlanckCollaboration2013,Greco2015,Ruan2015,Crichton2016}.

In tSZ-SPT and tSZ-ACT, they used simple models of heating due to gravitation and AGN feedback to compare with observations. For gravity, they assumed that as gas collapses and virializes along with an encompassing spherical halo of dark matter, it is shock-heated to a virial temperature $T_{\rm vir}$. For isothermal gas, this gives a total thermal energy of
\begin{equation}
\begin{split}
E_{\rm therm,gravity} &= \frac{3 k T_{\rm vir}}{2} \frac{\Omega_b}{\Omega_0} \frac{M}{\mu m_p}\\ &=
1.5 \times 10^{60} \, {\rm erg} \, M_{13}^{5/3} (1+z),
\end{split}
\label{eq:simEtherm}
\end{equation}
where $m_p$ is the proton mass, $\mu = 0.62$ is the average particle mass in units of $m_p$, and $M_{13}$ is the mass of the halo in units of $10^{13} M_\odot.$ The relation between halo mass and stellar mass is derived from \citet{Ferrarese2002} and \citet{Marconi2003} and given by $M_{\rm stellar} 
 = 2.8_{-1.4}^{+2.4}  \times 10^{10} M_\odot   \,  M_{13}^{5/3} (1+z)^{5/2}.$ Incorporating this gives
\begin{equation}
\begin{split}
&E_{\rm therm,gravity} = \\ &5.4_{-2.9}^{+5.4} \times 10^{60} \, {\rm erg} \,  \frac{M_{\rm stellar}}{10^{11} M_\odot} (1+z)^{-3/2}.
\end{split}
\label{eq:simEgrav}
\end{equation}
This is the total thermal energy expected around a galaxy of stellar mass $M_{\rm stellar}$ ignoring both radiative cooling, which will decrease $E_{\rm therm},$ and AGN feedback, which will increase it.

\begin{figure*}[t]
\centerline{\includegraphics[height=7cm]{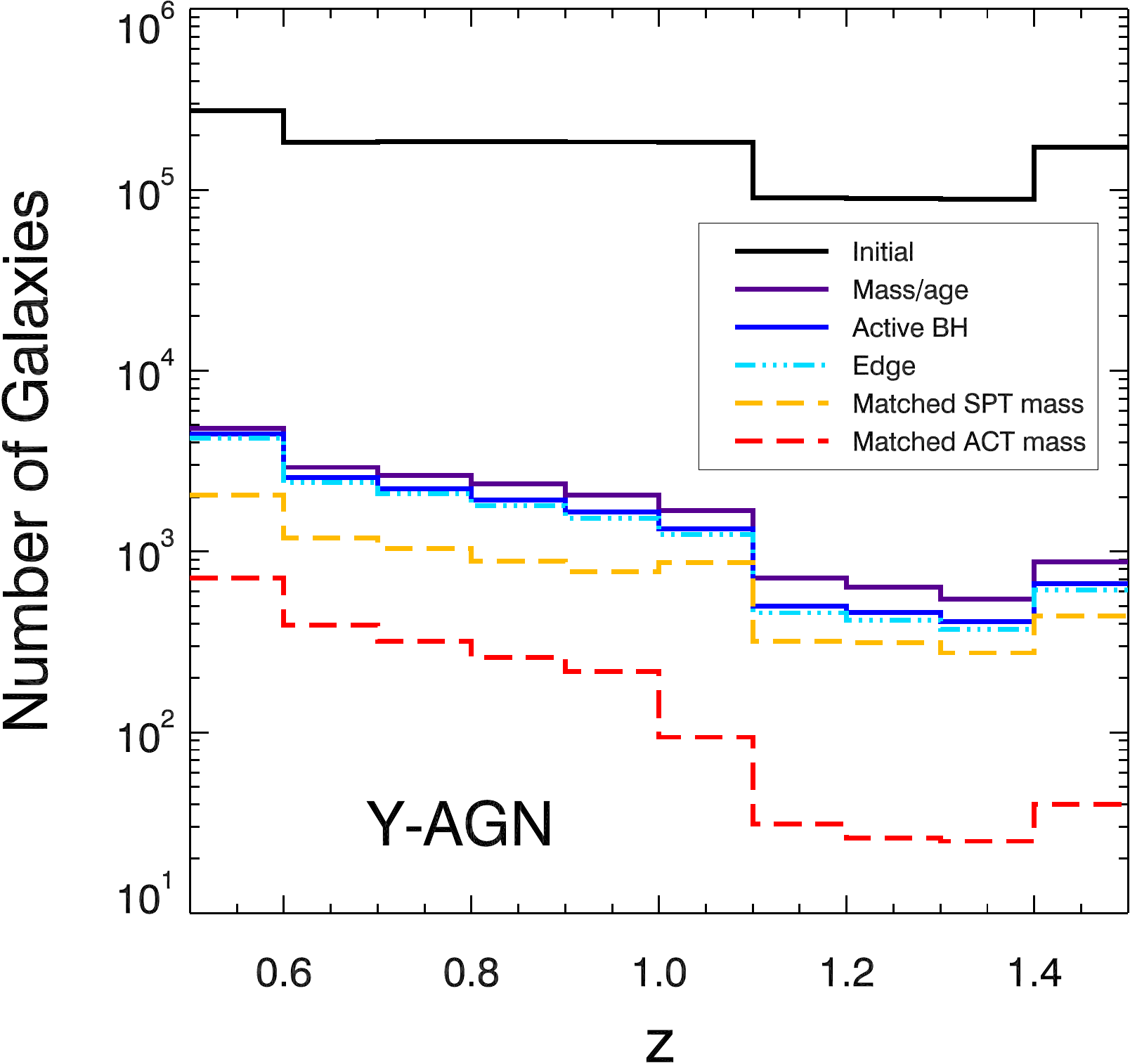}
\includegraphics[height=7cm]{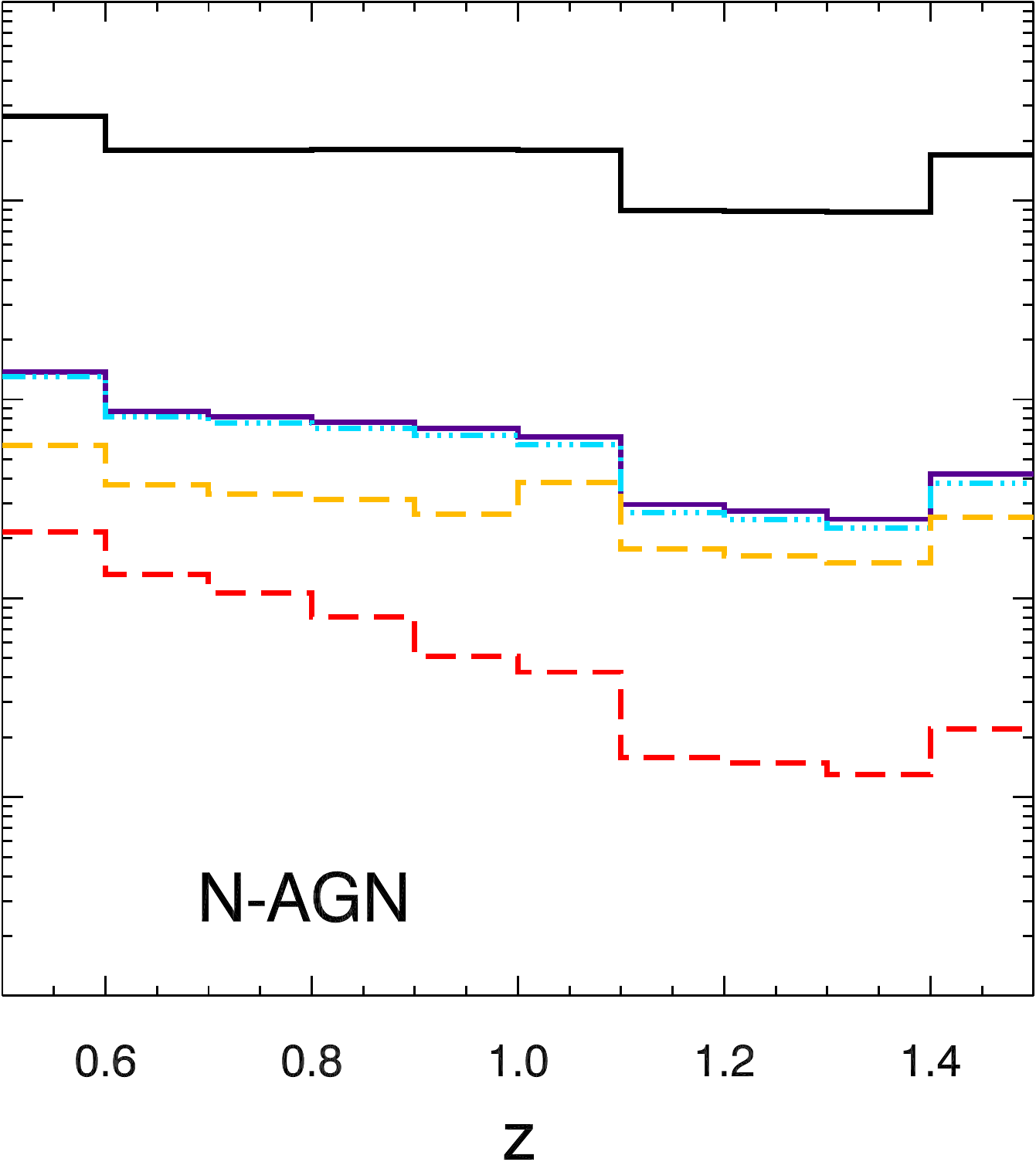}}
\caption{Number of galaxies per redshift bin for the initial population (black), after applying mass and age limits (purple), after removing active black holes (blue), after removing galaxies near the simulation edge (light blue dot-dashed), and then after matching both the tSZ-SPT (orange dashed) and tSZ-ACT (red dashed) mass distributions. On the left is the Y-AGN simulation and on the right is the N-AGN simulation.}\vspace{2mm}
\label{fig:simdata}
\end{figure*}

\begin{figure*}[t]
\centerline{\includegraphics[height=6.5cm]{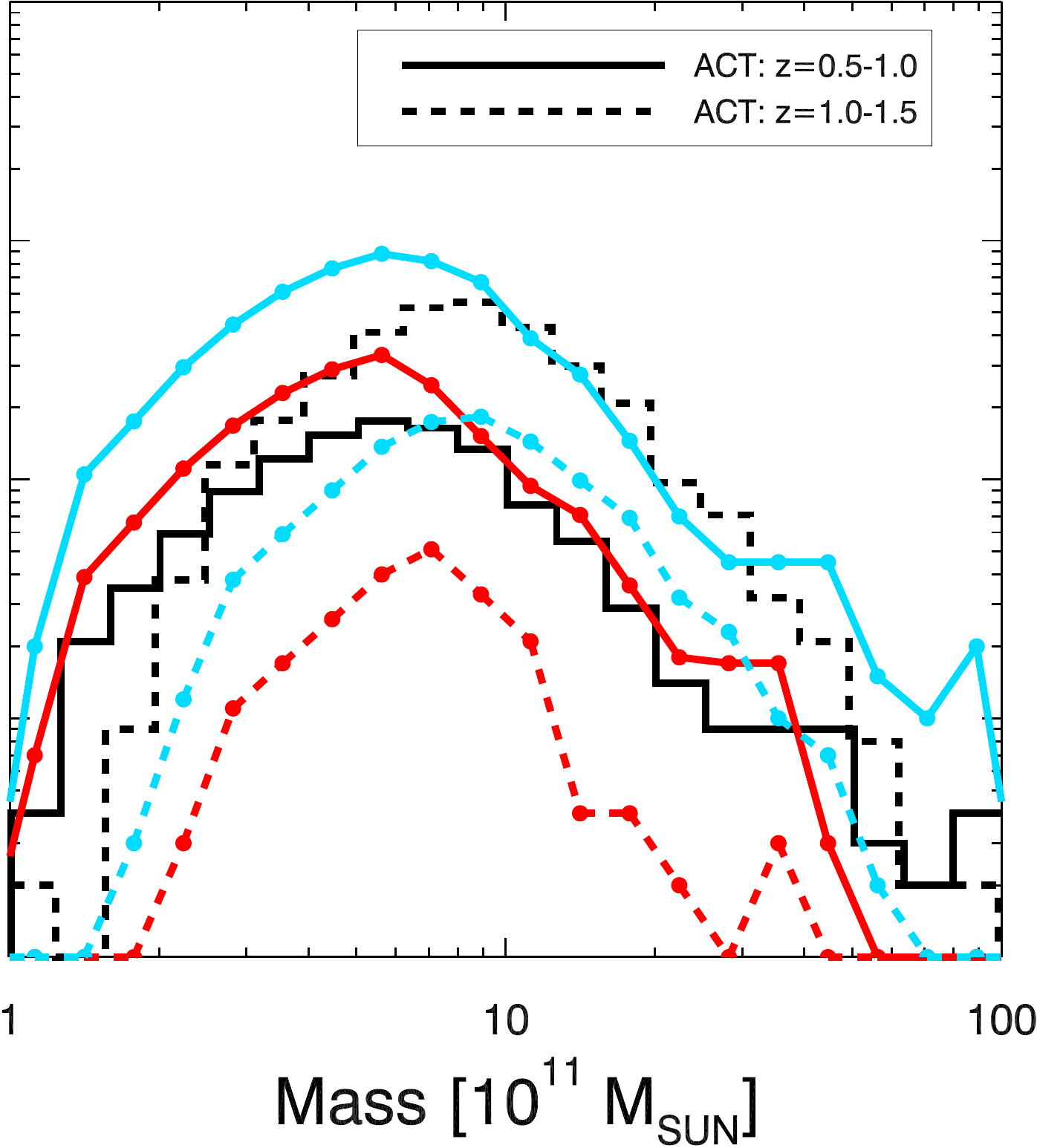}
\includegraphics[height=6.5cm]{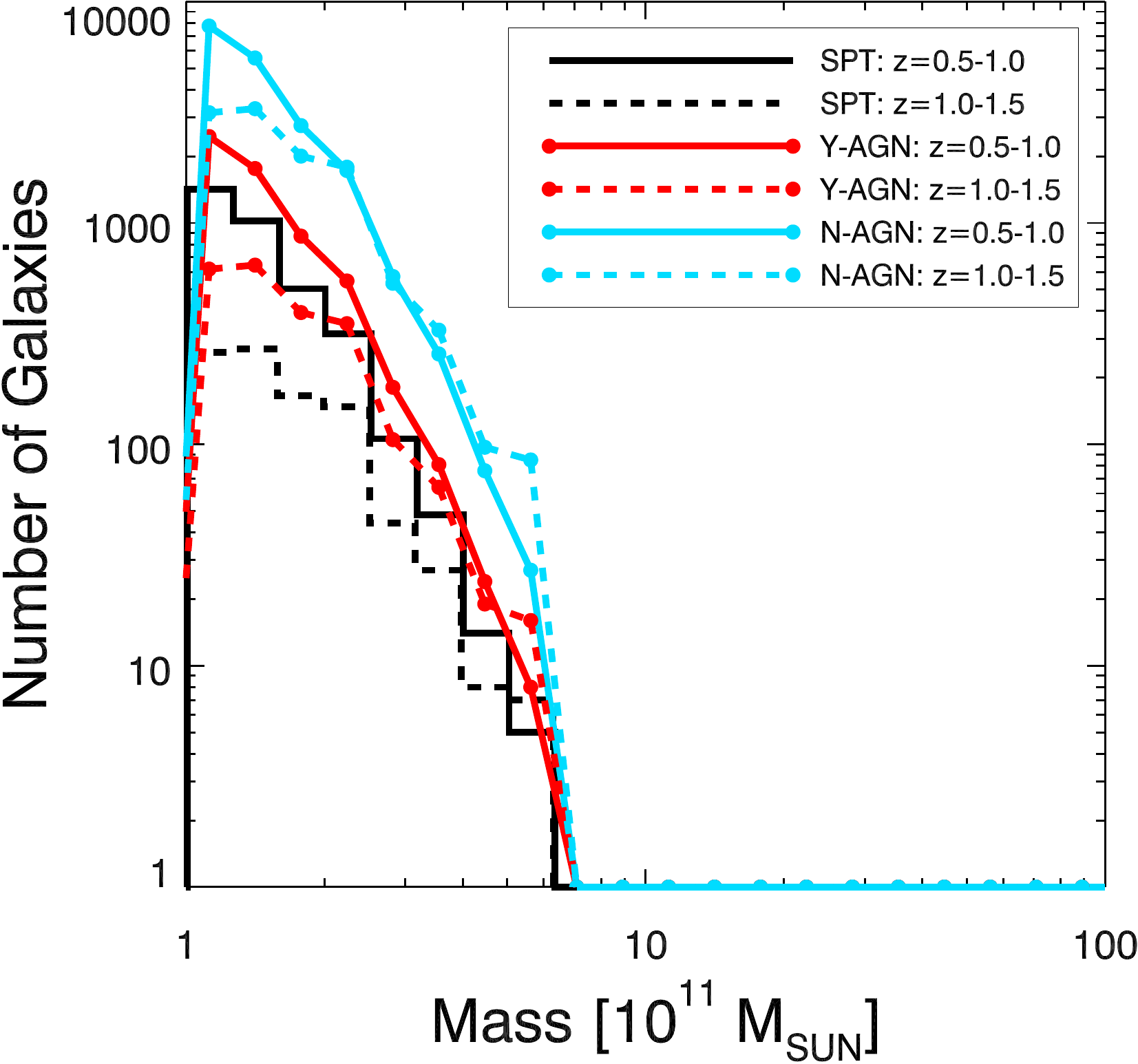}}
\caption{Mass selection comparisons. Solid lines represent the low-$z$ sample and dashed lines represent the high-$z$ sample. In the left plot, black represents the tSZ-SPT distribution, red represents the Y-AGN sample matched to tSZ-SPT, and light blue represents the N-AGN sample matched to tSZ-SPT. In the right plot, black represents the tSZ-ACT distribution, red represents the Y-AGN sample matched to tSZ-ACT, and light blue represents the N-AGN sample matched to tSZ-ACT.} \vspace{2mm}
\label{fig:matchmass}
\end{figure*}

For AGN feedback, they used the simple model described in \citet{Scannapieco2004}, where the black hole emits energy at the Eddington luminosity, $L_\text{Edd} = 1.26 \times 10^{38} (M_\text{BH}/M_\odot) \text{erg}\,\, \text{s}^{-1}$ \citep[e.g.][]{Shankar2013}, for a time $0.035 \, t_{\rm dynamical}$, with $t_{\rm dynamical} \equiv R_{\rm vir}/v_c = 2.6 \, {\rm Gyr} \, (1+z)^{-3/2}$ where $R_{\rm vir}$ is the virial radius and $v_c$ is the circular velocity of the galactic halo. The gas is heated by a fraction $\epsilon_{k}$ of the total bolometric luminosity of the AGN. This gives
\begin{equation}
\begin{split}
&E_{\rm therm, feedback} = \\ &4.1 \times 10^{60} \, {\rm erg} \,  \epsilon_{k,0.05}  \,  \frac{M_{\rm stellar}}{10^{11} M_\odot} \, (1+z)^{-3/2},
\end{split}
\label{eq:simEAGN}
\end{equation}
where $\epsilon_{k,0.05} \equiv \epsilon_{k}/0.05$ represents a typical efficiency factor of 5\%.


\section{The Horizon Simulations}
\label{sec:simhorizon}

\begin{figure*}[t]
\centerline{\includegraphics[height=6.5cm]{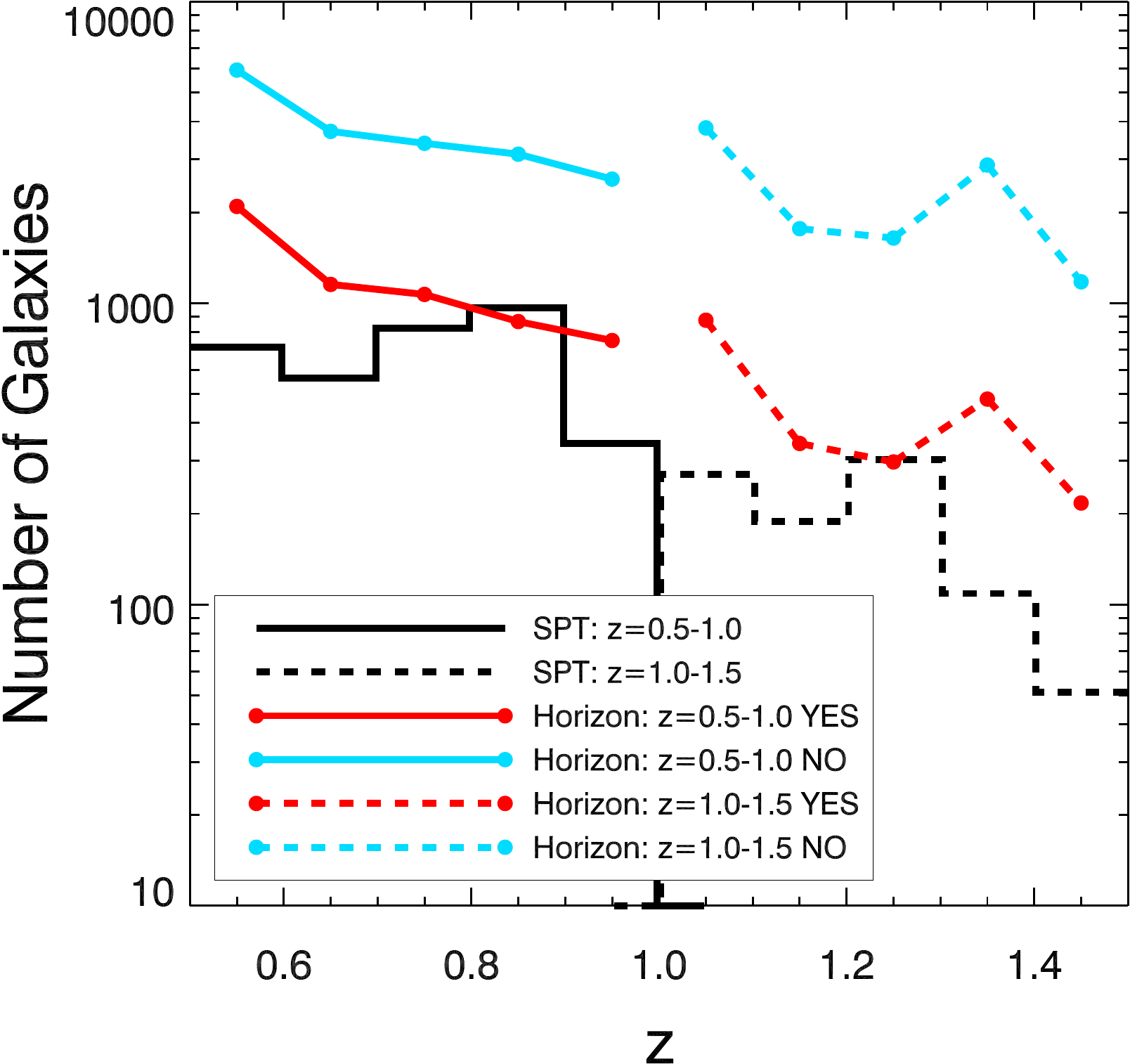}
\includegraphics[height=6.5cm]{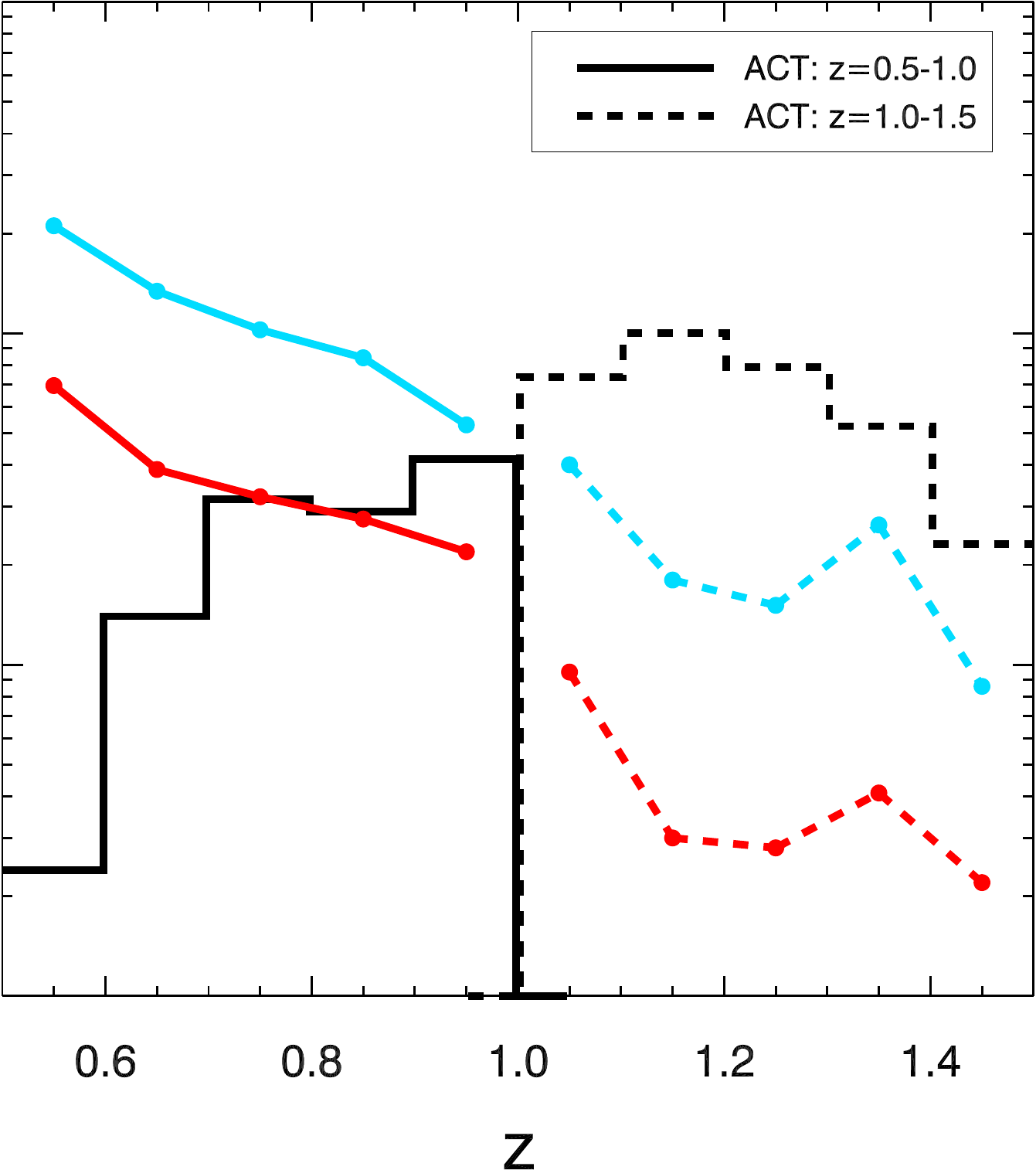}}\vspace{2.2mm}
\centerline{\includegraphics[height=6.5cm]{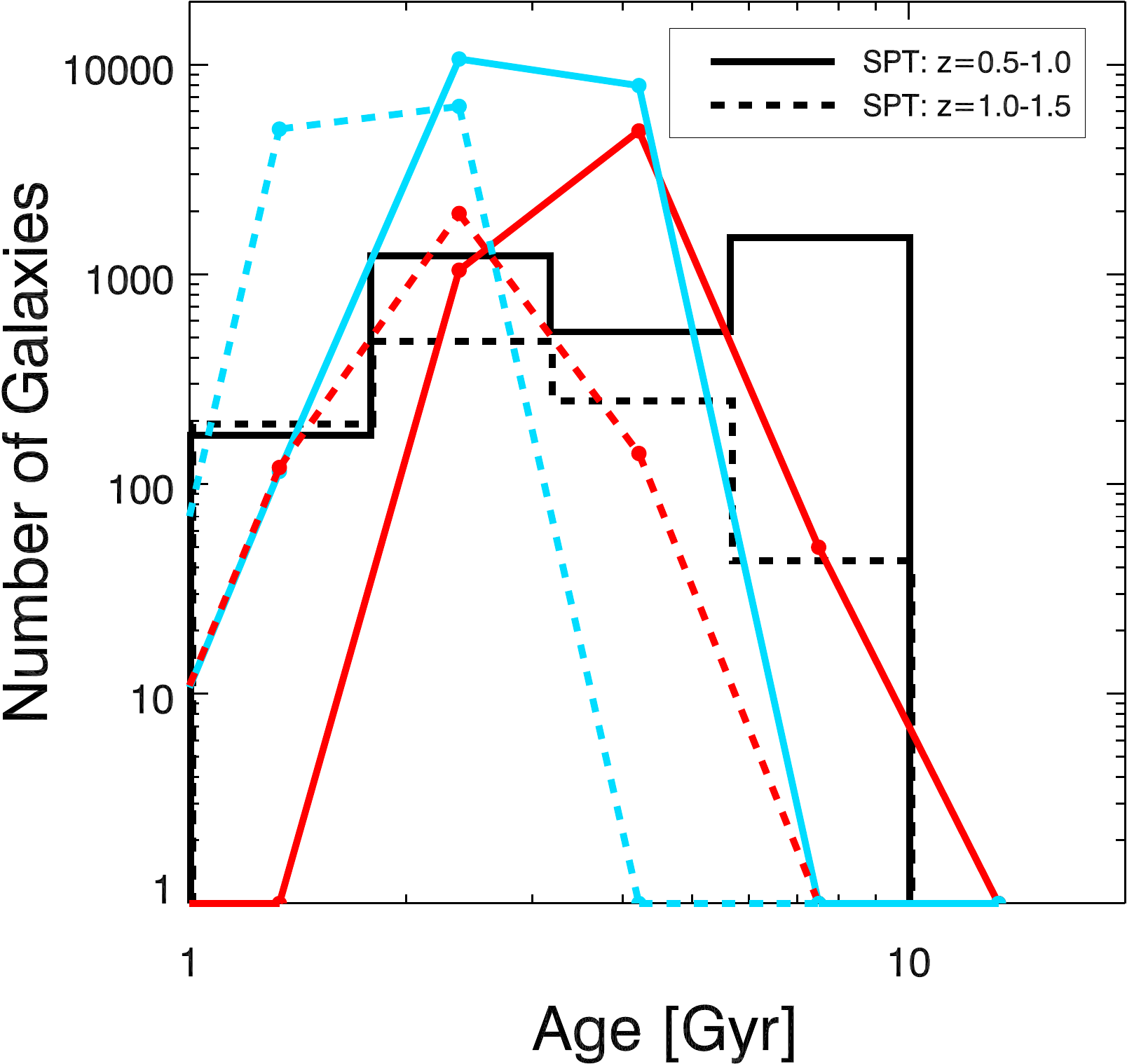}
\includegraphics[height=6.5cm]{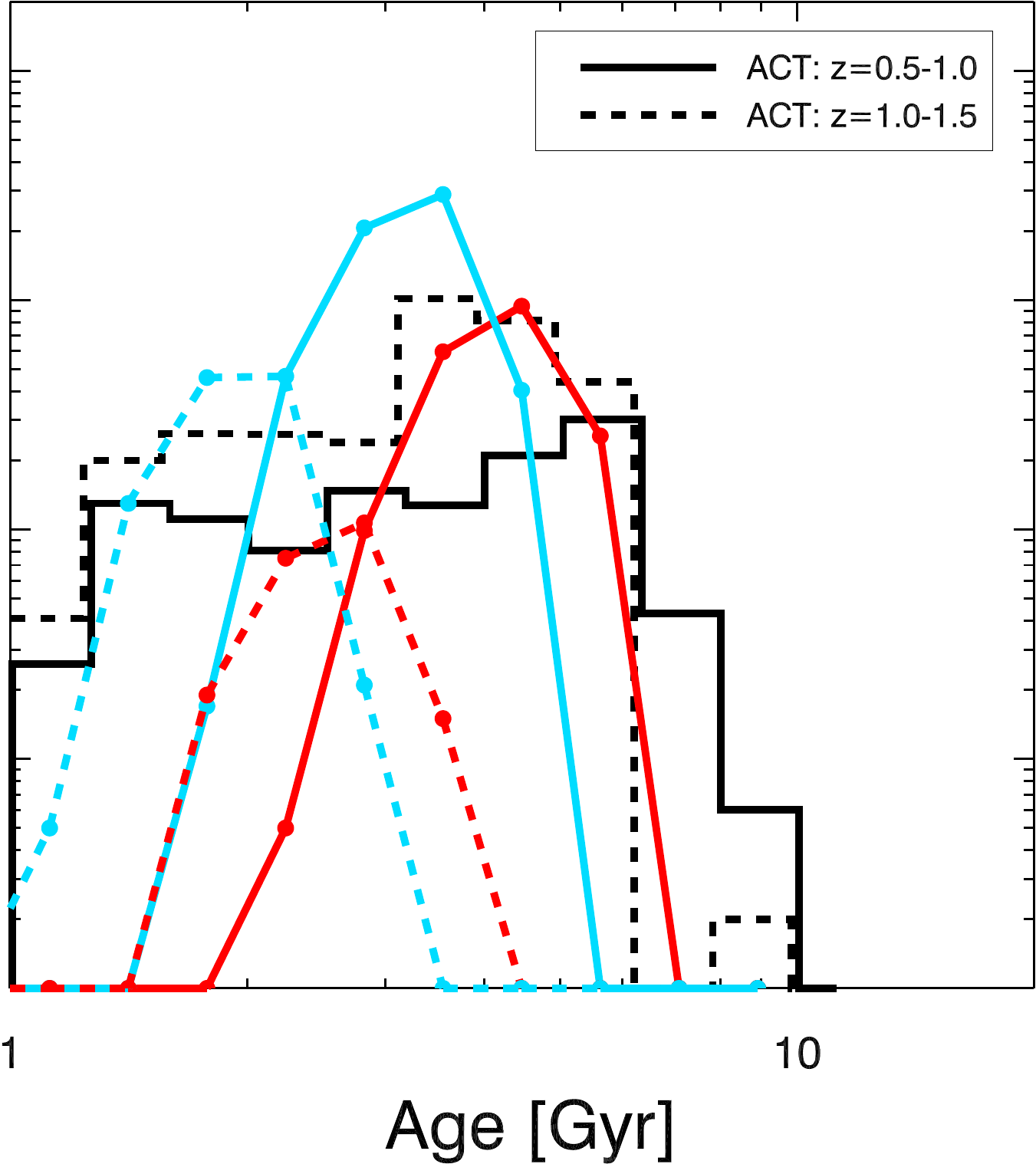}}
\caption{Redshift and age selection comparisons. The left plots are comparisons with SPT and the right plots are comparisons with ACT. The lines are the same as in Figure \ref{fig:matchmass}.\vspace{2mm}}
\label{fig:matchzage}
\end{figure*}

The Horizon-AGN simulation \citep{Dubois2014} is a cosmological hydrodynamical simulation that uses the adaptive mesh refinement Eulerian hydrodynamics code RAMSES \citep{Teyssier2002}. A complementary Horizon-noAGN simulation has also been performed that is identical to Horizon-AGN but has no prescription for black hole feedback \citep{Peirani2017}. The simulation box is 100/$h$ Mpc comoving on each side, with $1024^3$ dark matter particles at a mass resolution of $8 \times 10^7 M_\odot$. The simulation grid is refined throughout the simulation, with a maximum cell resolution of 1 physical kpc. Gas cools through emission by H, He, and metals \citep{Sutherland1993}, and is heated by a uniform UV background \citep{Haardt1996}. Stars are created following a Poissonian random process \citep{Dubois2008} in gas denser than a hydrogen number density of 0.1 cm$^{-3}$ following the Schmidt law, $\dot{\rho}_* = \epsilon_*\rho/t_{\rm ff}$, where $\dot{\rho}_*$ is the star formation rate density, $\epsilon_*$ = 0.02 is the star formation efficiency \citep{Krumholz2007}, and the gas with density $\rho$ has free-fall time $t_{\rm ff}=\sqrt{3\pi/(32G\rho)}$ where $G$ is the gravitational constant. The simulation uses a star particle mass resolution of $\approx 2 \times 10^6 M_\odot$. Stellar feedback is incorporated assuming a \citet{Salpeter1955} initial mass function, with stellar winds and mechanical energy from Type II supernovae taken from STARBURST99 \citep{Leitherer1999} and Type Ia supernova frequency taken from \citet{Greggio1983}.

\begin{table}[t]
\begin{center}
\resizebox{8.5cm}{!}{
\begin{tabular}{|c|c|c|c|} \hline
 Final bin: & Number & N(tSZ-SPT) & N(tSZ-ACT) \\ \hline
 SPT low-$z$ Y-AGN: &         5932 & \multirow{2}{*}{937} & \multirow{2}{*}{-} \\
 SPT low-$z$ N-AGN: &        18724 &   &   \\ \hline
 SPT high-$z$ Y-AGN: &         2213 & \multirow{2}{*}{240} & \multirow{2}{*}{-} \\
 SPT high-$z$ N-AGN: &        11265 &   &   \\ \hline
 ACT low-$z$ Y-AGN: &         1898 & \multirow{2}{*}{-} & \multirow{2}{*}{227} \\
 ACT low-$z$ N-AGN: &         5845 &   &   \\ \hline
 ACT high-$z$ Y-AGN: &          216 & \multirow{2}{*}{-} & \multirow{2}{*}{529} \\
 ACT high-$z$ N-AGN: &         1082 &   &   \\ \hline
 \end{tabular} }
\end{center}
\caption{Final number of galaxies in each redshift bin (``low-$z$'' = $0.5 < z < 1.0$ and ``high-$z$'' = $1.0 < z < 1.5$) for each corresponding survey, compared with the numbers from tSZ-SPT and tSZ-ACT that included Planck contamination modeling. \vspace{2mm}}
\label{tab:finnums}
\end{table}

\begin{table*}[t]
\begin{center}
\resizebox{14cm}{!}{
\begin{tabular}{|c|c|c|c|c|c|c|c|c|} \hline
 Survey & $z$ & $\left< z \right>$ & $\left< l_\text{ang}^2 \right>$ & $\left< M \right>$ & $\left< \text{Age} \right>$ & $\left< z \right>_M$ & $\left< l_\text{ang}^2 \right>_M $ & $Y$ \\
  &  &  & (Gpc$^2$) & ($10^{11} M_\odot$) & (Gyr) &  & (Gpc$^2$) & ($10^{-7} \text{Mpc}^2$) \\ \hline
 &&&&&&&&\\[-1em]
 SPT & 0.5-1.0 & 0.72 & 2.30 & 1.51 & 4.34 & 0.72 & 2.30 & $2.3_{-0.7}^{+0.9}$ \\[2pt]
 Y-AGN SPT & 0.5-1.0 & 0.70 & 2.15 & 1.52 & 3.88 & 0.70 & 2.15 & $ 2.6_{- 0.3}^{+17.9}$ \\[2pt]
 N-AGN SPT & 0.5-1.0 & 0.71 & 2.18 & 1.51 & 3.05 & 0.70 & 2.17 & $ 1.1_{- 0.1}^{+ 4.6}$ \\ \hline
  &&&&&&&&\\[-1em]
 SPT & 1.0-1.5 & 1.17 & 3.02 & 1.78 & 2.64 & 1.19 & 3.03 & $1.9_{-2.0}^{+2.4}$ \\[2pt]
 Y-AGN SPT & 1.0-1.5 & 1.20 & 2.98 & 1.72 & 2.47 & 1.20 & 2.98 & $ 3.1_{- 0.4}^{+ 7.7}$ \\[2pt]
 N-AGN SPT & 1.0-1.5 & 1.22 & 2.99 & 1.71 & 1.85 & 1.22 & 2.99 & $ 0.9_{- 0.0}^{+ 2.0}$ \\ \hline
  &&&&&&&&\\[-1em]
 ACT & 0.5-1.0 & 0.83 & 2.56 & 7.81 & 3.80 & 0.86 & 2.61 & $1.2_{-1.4}^{+1.4}$ \\[2pt]
 Y-AGN ACT & 0.5-1.0 & 0.69 & 2.14 & 6.46 & 4.22 & 0.69 & 2.13 & $ 9.4_{- 2.0}^{+23.0}$ \\[2pt]
 N-AGN ACT & 0.5-1.0 & 0.68 & 2.13 & 7.73 & 3.24 & 0.69 & 2.13 & $ 2.0_{- 0.2}^{+ 6.6}$ \\ \hline
 \ &&&&&&&&\\[-1em]
 ACT & 1.0-1.5 & 1.20 & 3.04 & 10.1 & 3.56 & 1.21 & 3.05 & $1.9_{-1.0}^{+1.1}$ \\[2pt]
 Y-AGN ACT & 1.0-1.5 & 1.20 & 2.98 & 7.53 & 2.57 & 1.19 & 2.97 & $16.8_{- 6.1}^{+66.9}$ \\[2pt]
 N-AGN ACT & 1.0-1.5 & 1.21 & 2.99 &  9.85 & 1.95 & 1.20 & 2.98 & $ 3.3_{- 0.5}^{+ 5.4}$ \\ \hline
\end{tabular} }
\end{center}
\caption{Mean ($\left< x \right>$) and mass-averaged ($\left< x \right>_M$) values for several relevant galaxy parameters, comparing tSZ-SPT and tSZ-ACT with the matched Horizon galaxies. \vspace{2mm}}
\label{tab:avgs}
\end{table*}

AGN feedback in the Horizon-AGN simulation is modeled following \citet{Dubois2012}. Black holes are seeded at $10^5 M_\odot$ when the gas mass density $\rho > 0.1\; m_H/\text{cm}^3$, and the black hole accretion rate follows a Bondi-Hoyle-Lyttleton rate with an $\alpha$ boost factor \citep{Booth2009} that scales as $\rho^2$ for $\rho > 0.1\; m_H/\text{cm}^3$ and is 1 otherwise, accounting for high density gas that is not resolved in the simulation. The maximum accretion rate is set by the Eddington limit with a radiative efficiency of 0.1 \citep{Shakura1973}. When the accretion rate is high \citep[$> 0.01 L_\text{Edd}$, following][]{Merloni2008}, a quasar-like feedback mode is assumed with 1.5\% of the accretion energy injected as thermal energy in the surrounding gas \citep[e.g.][]{Booth2009}. When the accretion rate is low ($< 0.01 L_\text{Edd}$), a radio feedback mode is assumed with bipolar jets at 10\% efficiency, calibrated in \citet{Dubois2012}. Feedback parameters are chosen to match the $M_\text{BH}-M_\star$ \citep{Haring2004} and $M_\text{BH}-\sigma_\star$ \citep{Tremaine2002} relations observed at $z = 0$. Black hole sink particles merge when they are closer than 4 kpc and slower than their mutual escape velocity. Black holes are free to leave the center of their galaxies, although they are subject to a physically motivated gas dynamical drag force term, $F_\text{drag} = f_\text{gas} 4\pi \alpha \rho (G M_\text{BH}/\bar{c}_s)^2$, where $f_\text{gas}$ is a fudge factor between 0 and 2, $\alpha$ is the same as above, and $\bar{c}_s$ is the average sound speed \citep{Volonteri2016}. This acts as a dynamical friction term, exerted by the gas on to the black hole \citep{Dubois2013}, and it should have the effect of keeping black holes in the center of their galaxies. A mock observational image of the Horizon-AGN simulation results showing a subset of the galaxies used in this work (green circles) is shown in Figure \ref{fig:examplegal}.


\section{Data}
\label{sec:simdata}

We use the Horizon-AGN (abbreviated hereafter as ``Y-AGN'') and Horizon-noAGN (abbreviated hereafter as ``N-AGN'') simulations for our simulated galaxy data, and make comparisons with observational measurements from tSZ-SPT and tSZ-ACT. In order to obtain as robust of a galaxy sample as possible, we collect data from the full available spread of 18 redshift outputs between z=0.5 and 1.5. We initially find all galaxies in the simulations with at least 250 star particles at each redshift. We then extract various parameters for each galaxy (stellar mass, age, star formation rate, and active black hole flag in the Y-AGN case) and corresponding dark matter halo (total mass, gas mass, stellar mass, black hole mass, and dark matter mass) in both Y-AGN and N-AGN. The active black hole flag says that a black hole is active if its mass is greater than $10^6 M_\odot$ and its luminosity is greater than $0.01L_\text{Edd}$. The stellar mass is taken within a virial radius $r_{200}$ of a galaxy, defined as the radius enclosing an overdensity of 200 times the critical density, given by
\begin{equation} \frac{M(<r_{200})}{\frac{4}{3}\pi r_{200}^3} = 200 \rho_\text{c}(z) ,\end{equation}
where $M(<r_{200})$ is the mass enclosed within $r_{200}$, and $\rho_\text{c}(z) = 3H^2(z)/(8\pi G)$.

We then perform a cut on the galaxies corresponding to the selection criteria in tSZ-SPT and tSZ-ACT, requiring $M_\text{stellar} \geq 10^{11} M_\odot$, Age $\geq 1$ Gyr, and a non-active black hole in the Y-AGN case. Although the simulation is periodic, we omit those few galaxies within 4 arcmin of the box edge to limit the computational overhead for correctly extracting the tSZ signal on the scales of ACT and SPT. The number of galaxies remaining in each redshift bin after the initial parameter cuts and the active black hole cut is shown in Figure \ref{fig:simdata}. Note that we see a steady decrease in the fraction of galaxies that are flagged as active as we decrease in redshift, generally $> 20\%$ for $1.0 < z < 1.5$ (25\% mean) and $< 20\%$ for $0.5 < z < 1.0$ (15\% mean). This makes sense, since $z \approx 2$ represents the peak of AGN activity, and it should decrease to lower $z$ \citep{Volonteri2016}.

For each galaxy, we use the electron pressure to compute Compton-$y$ values (Equation (\ref{eq:simy})) and then project these values into two dimensions, representing the signal that would be observed. We then select a square region around each galaxy and simulate measurements corresponding to the beam and pixel sizes in both tSZ-SPT and tSZ-ACT. For tSZ-SPT, we use regions that are $8.25 \times 8.25$ arcmin, $33 \times 33$ pixels, convolved with a 1.15 arcmin FWHM Gaussian (corresponding to the 150 GHz SPT beam). For tSZ-ACT, we use regions that are $8.36 \times 8.36$ arcmin, $17 \times 17$ pixels, convolved with a 1.44 arcmin FWHM Gaussian (corresponding to the 148 GHz ACT beam). This results in a selection of galaxies for both Y-AGN and N-AGN at $0.5 < z < 1.5$ with tSZ measurements matching both tSZ-SPT and tSZ-ACT. 


\section{Measurements}
\label{sec:simmeasure}

\begin{figure*}[t]
\centerline{\includegraphics[height=6cm]{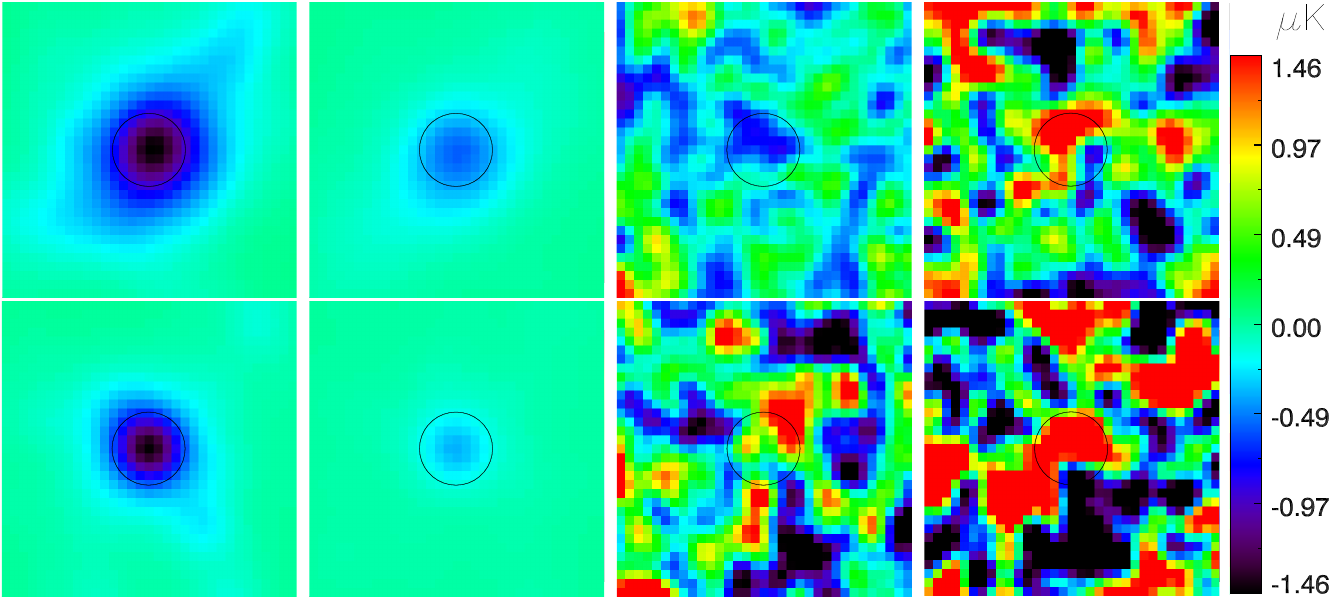}}
\caption{tSZ-SPT-matched 150 GHz stacked averages around galaxies for Y-AGN (left), N-AGN (left middle), low-$z$ (top), and high-$z$ (bottom). On the right are the initial tSZ-SPT stacking results with the same scale, for 150 GHz (right middle) and 220 GHz (right). Black circles represent a 1 arcmin radius. \vspace{2mm}}
\label{fig:SPTstamps}
\end{figure*}

\begin{figure*}[t]
\centerline{\includegraphics[height=6cm]{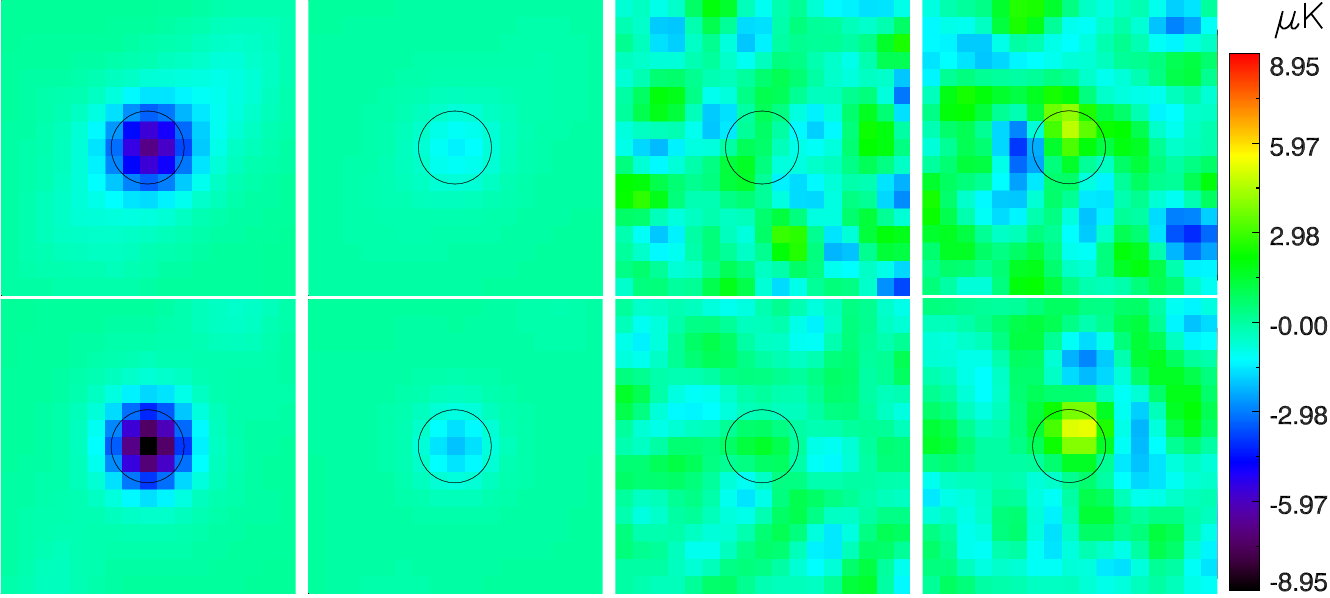}}
\caption{tSZ-ACT-matched 148 GHz stacked averages around galaxies for Y-AGN (left), N-AGN (left middle), low-$z$ (top), and high-$z$ (bottom). On the right are the initial tSZ-ACT stacking results with the same scale, for 148 GHz (right middle) and 220 GHz (right). Black circles represent a 1 arcmin radius. \vspace{2mm}}
\label{fig:ACTstamps}
\end{figure*}

In order to compare the simulated and observed galaxies, we randomly build a population of simulated galaxies until their mass distribution matches the histogram distributions in tSZ-SPT and tSZ-ACT. The matched distribution is scaled by whichever bin has the lowest fraction of Horizon galaxies compared to that bin for tSZ-SPT or tSZ-ACT. These matched mass distributions are shown in Figure \ref{fig:matchmass}, where the left plot shows the Y-AGN (red) and N-AGN (blue) galaxies matched to tSZ-SPT, and the right plot shows them matched to tSZ-ACT. Y-AGN and N-AGN galaxies are matched independently of one another since the matched properties can differ greatly between the two simulations. An issue arises since the tSZ-ACT mass distribution is skewed to higher masses while the overall Horizon distributions favor lower masses. This makes it difficult to match the tSZ-ACT distribution at the highest masses, especially for Y-AGN where there are fewer galaxies to begin with, so we choose a mass limit when necessary where we keep all Horizon galaxies above that mass. This is especially clear for the Y-AGN $z = 1.0-1.5$ line in the right plot of Figure \ref{fig:matchmass}, where we only match the tSZ-ACT distribution up to $\approx$ 10$^{12} M_\odot$ and keep every galaxy above that. In order to make sure that our results are insensitive to the random galaxy selection process, we run through several iterations of the process and find no significant change in the results.

The resulting tSZ-SPT, tSZ-ACT, and Horizon distributions for redshift and age are shown in Figure \ref{fig:matchzage}, the final number of Horizon galaxies at each redshift bin after this matching are shown in Figure \ref{fig:simdata}, and the final number of total galaxies in each redshift bin (``low-$z$'' is $0.5 < z < 1.0$ and ``high-$z$'' is $1.0 < z < 1.5$) for each survey is given in Table \ref{tab:finnums}. We note that the tSZ-SPT and tSZ-ACT numbers given in Table \ref{tab:finnums} represent the final measurements that incorporated Planck data, which used a smaller subset of galaxies than the original galaxy selection in tSZ-SPT and tSZ-ACT whose distributions are used in Figures \ref{fig:matchmass} and \ref{fig:matchzage}. The additional galaxy cuts for the final measurements using Planck data were only spatially dependent, so we expect the mass, redshift, and age distributions to be roughy the same as those distributions in Figures \ref{fig:matchmass} and \ref{fig:matchzage}. The tSZ measurement comparisons done in this work will use the Planck-incorporated tSZ-SPT and tSZ-ACT results.

The Horizon redshift distributions follow tSZ-SPT well, with the number of galaxies increasing with decreasing redshift. The Horizon redshift distributions do not follow tSZ-ACT as well, since tSZ-ACT peaks at higher masses (see Figure \ref{fig:matchmass}). Since mass tends to increase with decreasing redshift in the Horizon simulations, matching to the high tSZ-ACT masses results in a selection of lower redshift Horizon galaxies. The Horizon age distributions are similar to those from tSZ-SPT and tSZ-ACT, though N-AGN galaxies tend to be younger than Y-AGN galaxies because star formation proceeds in N-AGN while it is quenched by the feedback in Y-AGN \citep{Kaviraj2017}.

\begin{figure*}[t]
\centerline{\includegraphics[height=5.5cm]{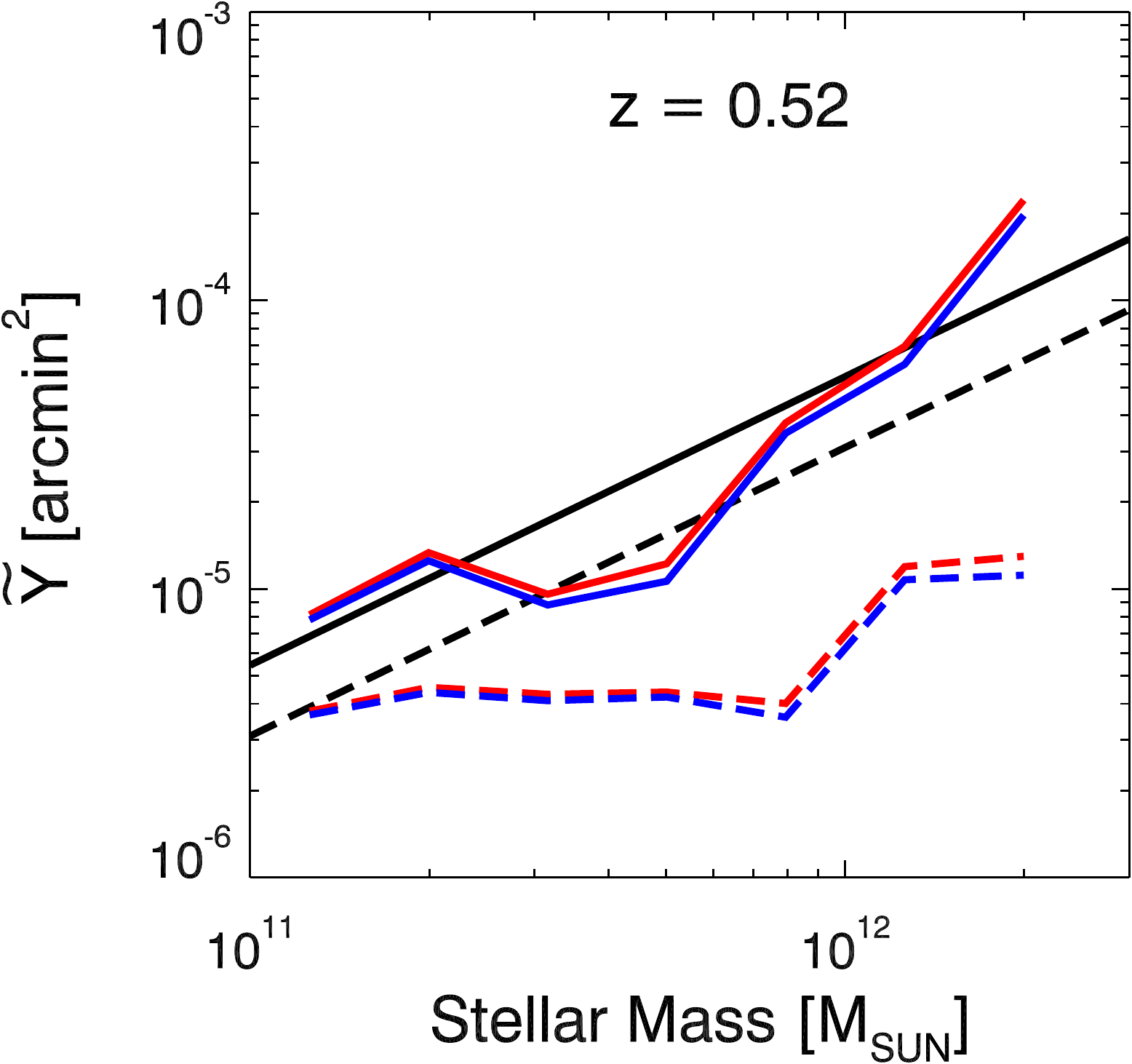}
\includegraphics[height=5.5cm]{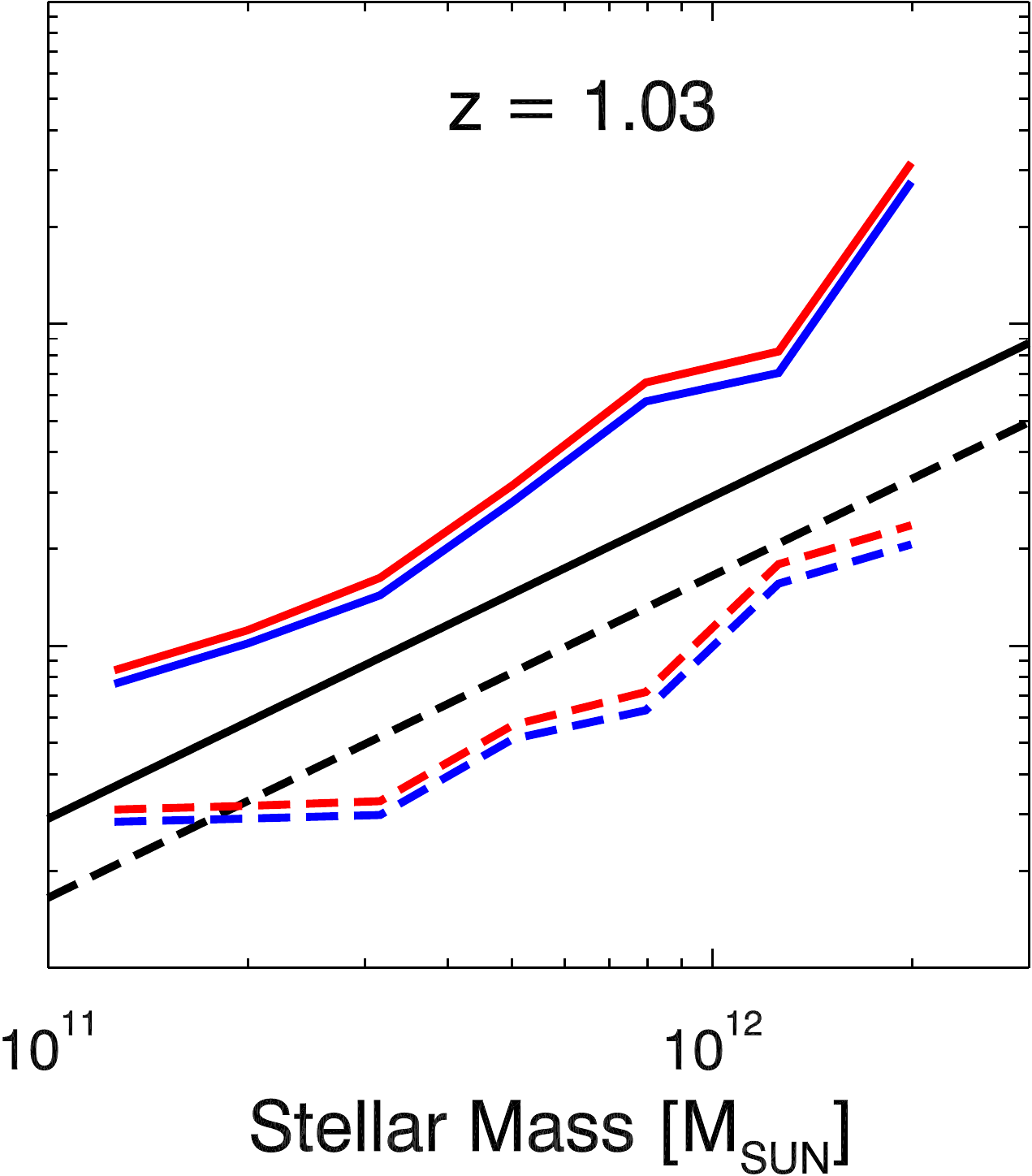}
\includegraphics[height=5.5cm]{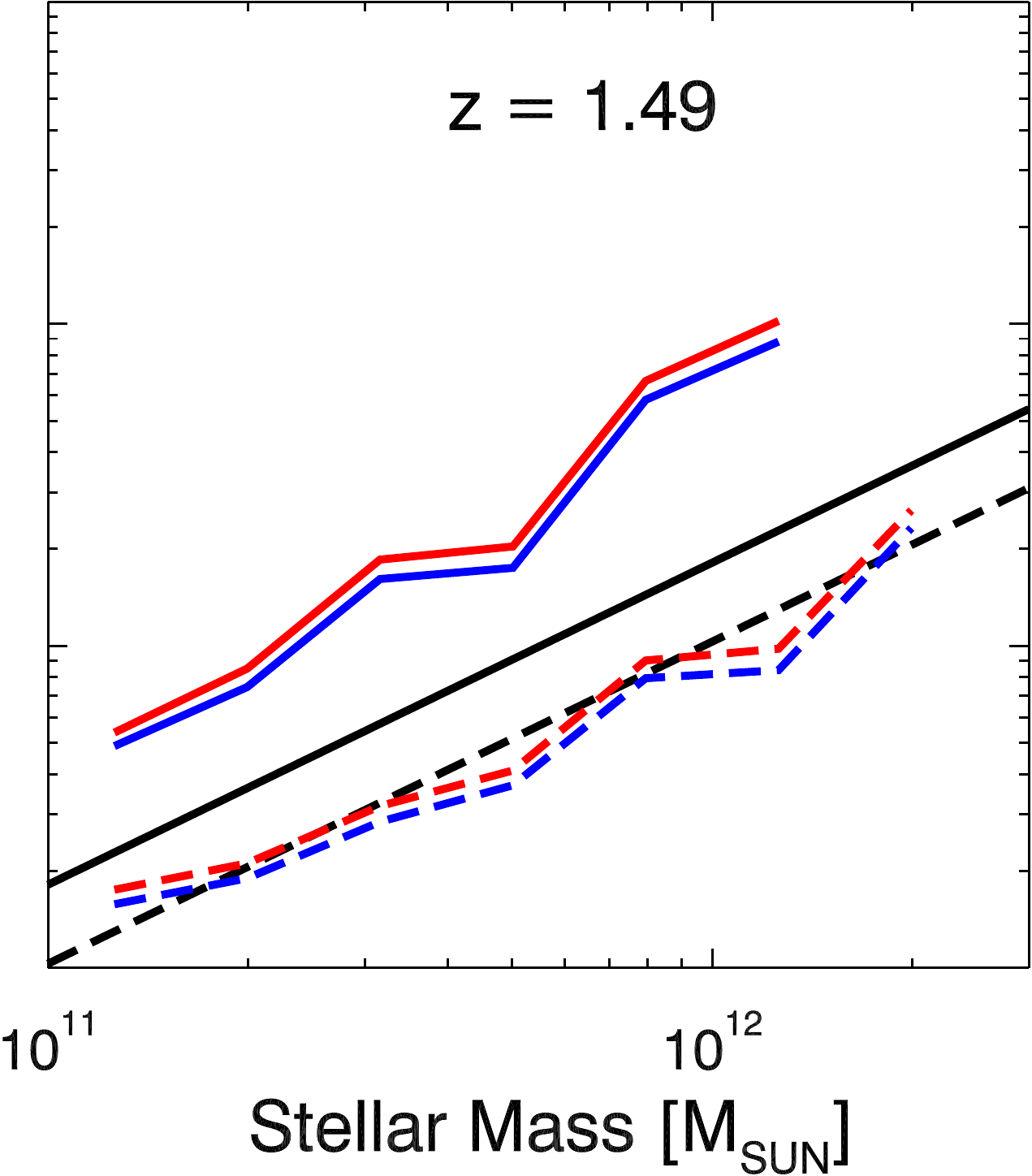}}
\caption{Comparison between scaled angularly-integrated, mass-binned Compton-$y$ measurements at a subset of simulation redshifts. Red lines are SPT-matched, blue are ACT-matched, solid are Y-AGN, and dashed are N-AGN. The solid black lines show Equation (\ref{eq:simEgrav}) + Equation (\ref{eq:simEAGN}) for the corresponding masses and redshifts, representing the simple AGN feedback energy model, and the dashed black lines show just Equation (\ref{eq:simEgrav}), representing only gravitational heating. \vspace{2mm}}
\label{fig:compare}
\end{figure*}

A comparison of averaged parameters from tSZ-SPT and tSZ-ACT with the final Horizon galaxies is shown in Table \ref{tab:avgs}. Notice that the average redshifts of the Horizon galaxies are very close to the corresponding tSZ-SPT and tSZ-ACT bins except for the tSZ-ACT low-$z$ bin. The tSZ-ACT low-$z$ distribution drops off towards $z = 0.5$ due to the selection method used, while the Horizon galaxies increase towards lower $z$ because only the mass distributions were matched, and higher mass galaxies tend to exist at lower redshifts. The average masses are all very similar except for the Y-AGN tSZ-ACT case, which struggles to match the high redshift distribution of tSZ-ACT with its limited number of galaxies. tSZ-SPT and tSZ-ACT galaxies tend to be older than the Horizon galaxies, again except for the Y-AGN tSZ-ACT case, which has older galaxies due to its greater number of low redshift galaxies compared to tSZ-ACT.

The errors in the simulation mean $Y$ values are computed based on finding the variance of the mean for skewed distributions, since for the individual galaxies there tend to be a lot of $Y$ values below the mean, but the values above the mean can reach high values and thus have a large range. This leads to much larger uncertainties above the mean $Y$ values than below. To compute these values we first take the mean $Y$ value, and separate the individual measurements based on whether they are below or above the mean. Next we account for the fact that the different simulation redshift outputs typically contain the same galaxies but at different times. When computing uncertainties, these same galaxies potentially being measured multiple times shouldn't improve the results, so we take a number $N_\text{ind}$ of independent galaxies for each group of galaxy measurements used. To estimate this number, we simply take the number of galaxies with $z = 0.72$ in the low-$z$ case and $z = 1.23$ in the high-$z$ case, representing the number of independent galaxies in a single redshift simulation output from roughly the middle of either redshift bin. For the galaxy measurements now separated to those below or above the mean, we compute the variance as
\begin{equation}
\text{var}  =  \frac{1}{N_\text{ind}} \sum_i{(Y_i-\left<Y\right>)^2},
\label{eq:Yvar}
\end{equation}
where $i$ runs over the $Y$ measurements and $\left<Y\right>$ is the mean $Y$ value. We then compute the uncertainty in the mean values by
\begin{equation}
\sigma  =  \sqrt{\frac{\text{var}}{N_\text{ind}}}.
\label{eq:Yerr}
\end{equation}

In addition to this statistical sample uncertainty there is also a potential effect from taking into account the noise in the original tSZ-SPT and tSZ-ACT measurements, but we note that this noise is not expected to have an effect on our results. The noise from the real data in tSZ-SPT and tSZ-ACT is estimated by making many stacks of random points in the sky. Assuming the noise fluctuations are unrelated to the signal being measured, if we were to add this noise to the the simulated signal in an additive way then it would just be the simulated measurements plus the same noise fluctuations we have measured with the same noise properties, as this process is linear. Therefore it wouldn't change anything.

When compared to low-$z$ tSZ-SPT, the Y-AGN $Y$ value is high by $0.4\sigma$, while the N-AGN value is low by $0.2\sigma$. For high-$z$ tSZ-SPT, the Y-AGN value is high by $0.6\sigma$ while the N-AGN value is low by $0.4\sigma$. When compared to low-$z$ tSZ-ACT, the Y-AGN value is high by $3.4\sigma$ while the N-AGN value is high by $0.5\sigma$. For high-$z$ tSZ-ACT, the Y-AGN value is high by $2.3\sigma$ while the N-AGN value is high by $1.4\sigma$. Given this, both the low-$z$ and high-$z$ tSZ-SPT results appear inconclusive. On the other hand, the low-$z$ tSZ-ACT results heavily disfavor the Y-AGN model of AGN feedback, while the high-$z$ tSZ-ACT results disfavor both the Y-AGN and N-AGN values, with the Y-AGN values more significantly different.

\begin{figure*}[t]
\centerline{\includegraphics[height=7cm]{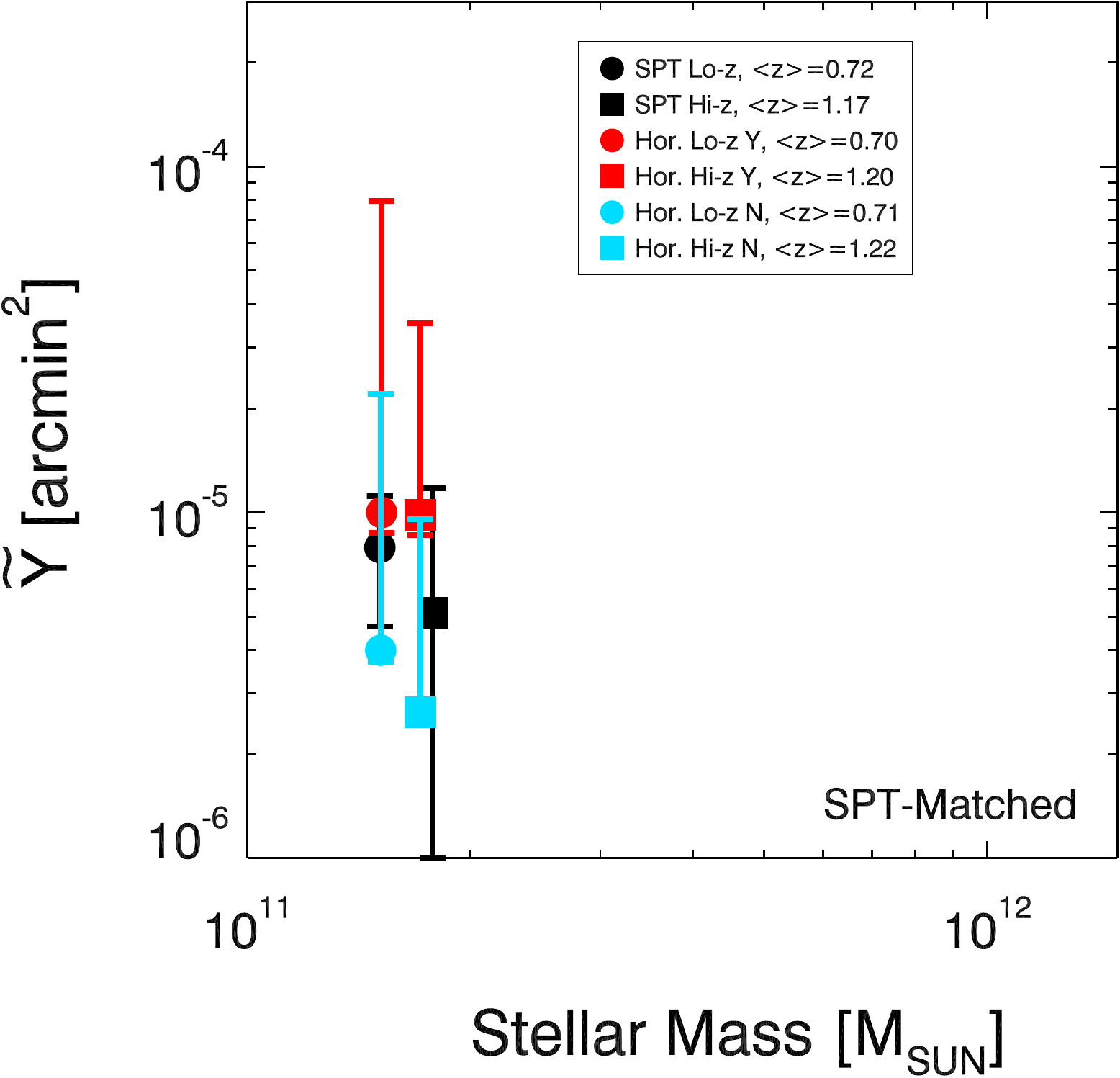}
\includegraphics[height=7cm]{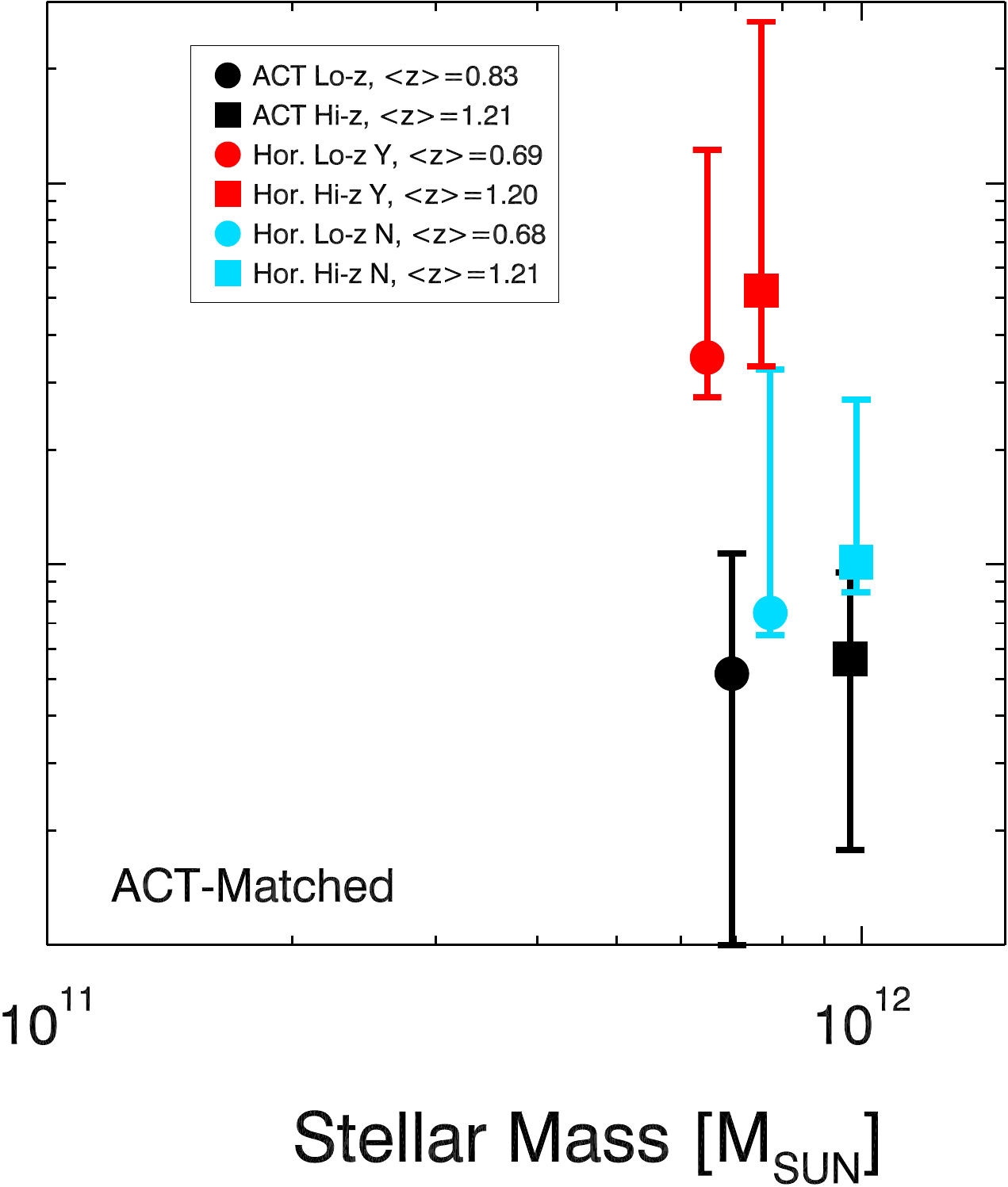}}
\caption{Final matched stack values for SPT and ACT. Red and blue circles are the total averaged low-$z$ values for Y-AGN and N-AGN, respectively, and red and blue squares are the total averaged high-$z$ values for Y-AGN and N-AGN, respectively. Black circles represent the tSZ-SPT or tSZ-ACT low-$z$ results, and black squares represent the tSZ-SPT or tSZ-ACT high-$z$ results. The left plot shows tSZ-SPT results with corresponding matched Horizon-AGN values, and the right plot shows tSZ-ACT results with the corresponding matched Horizon-AGN values. \vspace{2mm}}
\label{fig:fullplot}
\end{figure*}

\begin{table*}[t]
\begin{center}
\resizebox{15cm}{!}{
\begin{tabular}{|c|c|c|c|c|c|c|c|c|} \hline
 & \multicolumn{4}{c|}{SPT} & \multicolumn{4}{c|}{ACT} \\ \hline
 & \multicolumn{2}{c|}{low-$z$} & \multicolumn{2}{c|}{high-$z$} & \multicolumn{2}{c|}{low-$z$} & \multicolumn{2}{c|}{high-$z$} \\ \hline
Test & N-AGN & Y-AGN & N-AGN & Y-AGN & N-AGN & Y-AGN & N-AGN & Y-AGN \\ \hline
normal &  0.2 &  0.4 &  0.3 &  0.7 &  0.4 &  3.2 &  1.2 &  2.0 \\
alimit &  0.2 &  0.0 &  0.3 &  0.3 &  0.4 &  2.9 &  1.2 &  1.5 \\
blimit &  0.2 &  0.1 &  0.3 &  0.6 &  0.4 &  2.9 &  1.1 &  1.5 \\
dlimit &  0.2 &  0.5 &  0.3 &  0.6 &  0.3 &  3.2 &  1.2 &  2.7 \\
elimit &  0.2 &  0.6 &  0.3 &  0.7 &  0.3 &  3.1 &  1.0 &  1.8 \\
yproj &  0.3 &  0.1 &  0.3 &  0.5 &  0.2 &  3.1 &  1.1 &  2.0 \\
zproj &  0.3 &  0.1 &  0.3 &  0.4 &  0.2 &  3.1 &  1.1 &  2.5 \\
age0p6 &  0.2 &  0.4 &  0.3 &  0.6 &  0.4 &  3.1 &  1.1 &  2.5 \\
age1p5 &  0.2 &  0.5 &  0.3 &  0.6 &  0.3 &  3.3 &  1.0 &  2.1 \\
age1p8 &  0.2 &  0.3 &  0.2 &  0.7 &  0.3 &  3.2 &  1.1 &  2.0 \\
gdcut &  0.3 &  0.1 &  0.5 &  0.2 &  0.0 &  2.5 &  0.2 &  1.8 \\
upmcutp &  0.6 &  0.1 &  0.4 &  0.2 &  0.1 &  1.9 &  0.7 &  1.2 \\
upmcuts &  0.2 &  0.4 &  0.3 &  0.6 &  0.3 &  3.1 &  1.0 &  2.3 \\
offset0 &  0.2 &  0.4 &  0.3 &  0.7 &  0.4 &  2.9 &  1.2 &  2.1 \\
offset2 &  0.2 &  0.2 &  0.3 &  0.5 &  0.3 &  3.2 &  1.1 &  2.2 \\
maxmatch0 &  0.1 &  0.4 &  0.2 &  0.9 &  0.1 &  2.9 &  1.0 &  2.7 \\
ssfrcut1 &  0.1 &  0.4 &  0.2 &  0.6 &  0.1 &  1.4 &  1.0 &  0.6 \\
ssfrcut2 &  0.1 &  0.0 &  0.1 &  1.1 &  0.4 &  1.5 &  1.0 &  1.5 \\ \hline
MEAN &  0.2 &  0.3 &  0.3 &  0.6 &  0.3 &  2.8 &  1.0 &  1.9 \\
STDDEV &  0.1 &  0.2 &  0.1 &  0.2 &  0.1 &  0.6 &  0.2 &  0.5 \\ \hline
\end{tabular} }
\end{center}
\caption{Significance (number of $\sigma$, i.e. from Equation (\ref{eq:Yerr})) in the difference between the tSZ-SPT and tSZ-ACT $\tilde{Y}$ values and the results from this work, given for a wide variety of tests. See Section \ref{sec:simmeasure} for an explanation of the different tests. \vspace{2mm}}
\label{tab:tests}
\end{table*}

Final average stacked stamps for the tSZ-SPT redshift bins are shown in Figure \ref{fig:SPTstamps} for Y-AGN and N-AGN, along with the initial stamps from tSZ-SPT 150 and 220 GHz with matching scales. Final average stacked stamps for the tSZ-ACT redshift bins are shown in Figure \ref{fig:ACTstamps} for Y-AGN and N-AGN, along with the initial stamps from tSZ-ACT 148 and 220 GHz with matching scales. The scales for both figures are set by the highest and lowest pixels out of all four Horizon stamps. For tSZ-SPT, although the 220 GHz stamps are noisy, there is promise that the large 220 GHz signal, representing contamination, could produce a substantial tSZ decrement when removed from the 150 GHz stamps, especially at low-$z$ where there is already a decrement. For tSZ-ACT, it is clear that the tiny 220 GHz contaminants cannot account for the expected large tSZ decrement shown in the Y-AGN stamps. As shown in the tSZ-SPT and tSZ-ACT work, when these positive 220 GHz signals and the corresponding 150 GHz signals, along with additional measurements in the high-frequency Planck bands, are modeled as dust emission plus synchrotron emission plus tSZ signal, the most favorable models plus associated uncertainties reveal significant tSZ detections shown by the SPT and ACT $Y$ values in Table \ref{tab:avgs}.

The low-$z$ tSZ-SPT stamps also indicate a larger extent to the signal than for the high-$z$ stamps, $\approx 5$ arcmin compared to $\approx 3$ arcmin. At an average low-$z$ value of $z$=0.7, the ratio of physical size to angular size is 7.2 kpc/arcsec, while at an average high-$z$ value of $z$=1.2 it is 8.4 kpc/arcsec. This means the signal represents $\sim 1.5$ Mpc at high-$z$ and $\sim 2.2$ Mpc at low-$z$. The larger angular size of the signal at low-$z$ is therefore only slightly due to observational effects and mostly due to the signal being physically larger in size, possibly due to longer adiabatic expansion timescales (on average these are 1.5 times longer at low-$z$). This effect is not seen as clearly for tSZ-ACT, possibly due to the pixels being twice as large, or the masses, and therefore signal, being $\approx$ 6 times as large. These stamps also indicate that the 1 arcmin radius measurement aperture might be too small to contain the full tSZ signal. The low-$z$ tSZ-SPT stamps would have the full signal better contained within an aperture twice as large, while the high-$z$ tSZ-SPT stamps would need a 1.5-1.75 arcmin radius aperture. The tSZ-ACT stamps appear to need a 1.5-2 arcmin radius aperture to contain the full signal.

Mass binned averages of $\tilde{Y}$ for three redshifts spanning the full range comparing Horizon to the model predictions are shown in Figure \ref{fig:compare}. There are several trends to notice here. The best-fit (using $\chi^2$ analysis) linear log-log slopes and their associated 1$\sigma$ levels for $z=(0.52, 1.03, 1.49)$ are (1.1$\pm$0.2, 1.3$\pm$0.1, 1.3$\pm$0.1), respectively, for Y-AGN, and (0.4$\pm$0.1, 0.8$\pm$0.1, 0.8$\pm$0.1) for N-AGN. Thus the Y-AGN and N-AGN curves roughly follow the $\tilde{Y} \propto Y \propto E_\text{therm} \propto M_\star$ relation indicated by the simple models in Equations (\ref{eq:simEgrav}) and (\ref{eq:simEAGN}), except for the lowest redshift of $z = 0.52$. For the first four mass bins at $z = 0.52$ (representing $\approx 1-6\times10^{11} M_\odot$) the best-fit slopes are 0.2$\pm$0.2 for Y-AGN and 0.1$\pm$0.1 for N-AGN. For this lowest redshift bin, $\tilde{Y}$ appears to stay nearly flat for the lower masses. This decrease in the expected signal happens for both N-AGN and Y-AGN, but appears more prominent for N-AGN. The Y-AGN and N-AGN curves also stay roughly the same distance apart, with average vertical log distances between the Y-AGN and N-AGN lines of 0.7$\pm$0.2 at both $z=1.03$ and 1.49. At the highest redshift, the N-AGN line falls right on the model prediction with an average vertical difference of 0.02$\pm$0.09, while the Y-AGN line is considerably higher than the model prediction with an average vertical difference of 0.5$\pm$0.1. The simulated $\tilde{Y}$ values then both drop relative to the models as the redshift decreases, likely due to longer cooling times, which the models do not account for.

Final values of $\tilde{Y}$ for each redshift bin matched to each telescope, compared to the tSZ-SPT and tSZ-ACT results, are shown in Figure \ref{fig:fullplot}. Shown are the full averaged values with errors given by Equation (\ref{eq:Yerr}). As expected, the tSZ-SPT compared results appear to be mostly inconclusive, while the tSZ-ACT points seem to clearly match the N-AGN results much more than the Y-AGN results.

In order to test if we were introducing significant biases into our results based on our selection and measurement parameters, we performed a variety of different tests while varying a large number of these parameters. The results of these tests are shown in Table \ref{tab:tests}, where we give the significance as the number of $\sigma$ between the observation and simulation $\tilde{Y}$ results, i.e. the difference between the given results divided by the appropriate errors of the two results added in quadrature. ``Normal'' refers to the parameters used throughout this paper. ``Alimit,'' ``blimit,'' ``dlimit,'' and ``elimit'' refer to black hole Eddington luminosity fraction cuts of 0.001, 0.003, 0.03, and 0.1, respectively. ``Yproj'' and ``zproj'' refer to projecting our simulation galaxy measurements along the y-axis or z-axis of the simulation. ``Age0p6,'' ``age1p5,'' and ``age1p8'' refer to using galaxy age cuts of 0.6, 1.5, and 1.8 Gyr, respectively. ``Gdcut'' refers to removing all selected simulation galaxies within 1.5 arcmin of any other selected galaxies.

``Upmcutp'' and ``upmcuts'' refer to upper halo dark matter mass cuts applied to try and account for the galaxy cluster cuts used in tSZ-SPT and tSZ-ACT, where for ``upmcutp'' we have taken the limiting sensitivity of the Planck SZ cluster catalog \citep{PlanckCollaboration2016b} and converted it to a minimum halo mass at the average redshifts used here, and for ``upmcuts'' we have done the same for the SPT SZ cluster catalog \citep{Bleem2015}. To estimate the conversion from Compton-$y$ to halo mass, we take the relations between virial radius, virial temperature, halo mass, and redshift (i.e. equations 8 and 9 in tSZ-SPT), and combine them with Equation (\ref{eq:simy}) where we estimate $n_e/n_\text{tot} \approx 1 \times X (\mu/m_H) + 2 \times Y (\mu/m_{He}) \approx 1 \times 0.76 \times (0.62/(1 m_p)) + 2 \times 0.24 \times (0.62/(4 m_p))$, so that $n_e \approx 0.54 n_\text{tot}/(\mu m_p) \approx 0.88 \rho_\text{tot}/m_p \approx \rho_c(z)/m_p$. We also assume $T_e \approx T_\text{vir}$ and constant pressure so that the integral just becomes the size of the region of interest, which we take as $2 \times R_\text{vir}$. Putting it together, we get $T_\text{vir} \approx 2.2 \times 10^{10} \times y \times (1+z)$ K. We can then estimate halo mass with the relation $M_\text{halo} = 10^6 (T_\text{vir}/72\text{K})^{3/2} (1+z)^{-3/2} M_\odot$.

``Offset0'' and ``offset2'' refer to either not subtracting out the estimated background tSZ signal from the Horizon measurements, or subtracting out twice the computed offset. ``Maxmatch0'' refers to matching both the distribution and the exact numbers of galaxies in the tSZ-SPT and tSZ-ACT galaxy mass distributions, as much as possible. ``Ssfrcut1'' refers to applying the specific star formation rate (star formation rate divided by mass; SSFR) cut used in tSZ-SPT and tSZ-ACT, where SSFR $\leq 0.01 \text{Gyr}^{-1}$. ``Ssfrcut2'' refers to applying a more nuanced maximum SSFR cut following \citet{Liu2018}.

We also give the average values over all the tests and their spread. Generally, this reveals the inability of the tSZ-SPT comparisons to significantly distinguish between models with the current measurements and errors that we have. It also supports the clear indication that these results give towards a much smaller conflict between the observational results and the tSZ-ACT comparison N-AGN results, and a larger conflict with the tSZ-ACT comparison Y-AGN results especially at low-$z$. It is worth noting that the low values for ACT Y-AGN given by the ``ssfrcut1'' and ``ssfrcut2'' tests can be explained due to the sample having much lower masses (for high-$z$, only an average of $3.1 \times 10^{11} M_\odot$ in ``ssfrcut1'' and $3.3 \times 10^{11} M_\odot$ in ``ssfrcut2'', compared to $7.5 \times 10^{11} M_\odot$ in ``normal''), and tSZ signal decreases with decreasing mass as seen in Figure \ref{fig:compare}, as well as a much smaller sample size (for high-$z$, only 82 in ``ssfrcut1'' and 133 in ``ssfrcut2'', compared to 216 in ``normal''), leading to larger uncertainties.


\section{Discussion}
\label{sec:simdiscussion}

In this work, we have taken the large-scale cosmological hydrodynamical simulation Horizon-AGN, with AGN feedback, and its counterpart Horizon-NoAGN, without AGN feedback, and extracted a sample of galaxies that match those from tSZ-SPT and tSZ-ACT. We have then performed the same stacking procedure as tSZ-SPT and tSZ-ACT, measuring the tSZ effect by redshift, mass, and telescope survey. Comparing the simulation results to the observational results, we find that the tSZ-SPT results are consistent with both the Y-AGN and N-AGN results, while the tSZ-ACT results are consistent with the N-AGN curves and differ from the Y-AGN curves at $2-3\sigma$.

Overall, the tSZ-SPT comparison results fail to favor or disfavor the presence of AGN feedback, while the tSZ-ACT results indicate a tension between the simulation and observational measurements that suggests AGN feedback has less impact on the surrounding environment than expected. This impact by AGN feedback seems to be over-estimated by the Horizon-AGN simulation and likely other simulations that employ strong AGN feedback, at least for higher-mass ($\gtrsim 5 \times 10^{11} M_\odot$) galaxies (keeping in mind that AGN models vary significantly between different simulations). Simply reducing the efficiency of the feedback would not necessarily improve the agreement with tSZ-ACT, and would make the gas fraction in these halos even less consistent with other work, as seen in \citet{Chisari18}. Instead, more exploration of the implementation of feedback itself may improve the agreement of both our results and theirs, such as models that include solely momentum-driven feedback with jet-axis precession \citep[e.g.][]{Li2017,Weinberger2017}, cosmic-ray feedback \citep[e.g.][]{Sijacki2008}, the impact of magnetic fields on feedback \citep[e.g.][]{Dong2009}, and the different impacts radio-mode and quasar-mode feedback can have \citep[e.g.][]{Cielo2018}.

Another potentially important factor in simulations of the tSZ effect is the separation of electron and ion temperatures, which can be in nonequilibrium states and affect the measured tSZ Compton-$y$ signal by as much as 20\%, and the integrated $Y$ by as much as 9\% \citep{Rudd2009}. This bias would mean that the Horizon measurements are over-estimating the tSZ signal, and this would decrease the tension between the simulation and observation results, but we expect this effect to be small since it mostly affects the outskirts of clusters which we specifically try to avoid. Additionally, at large galaxy masses (especially at high redshift), where the conflict between observations and simulations appears largest, these simulations inevitably suffer from cosmic variance due to their limited size. Simulations with larger cosmological volumes may be needed to ensure that this is not only a sample bias effect. Still, these comparisons are unprecedented in their attempt to measure the tSZ effect around moderate redshift elliptical galaxies using the first publicly available data from the SPT and ACT surveys, and there remain many open questions that should be investigated regarding these tSZ stacking measurements.

Another notable observation from our results is that low-$z$ galaxies (i.e. $0.5 < z < 1.0$) appear to have a tSZ signal with significantly more angular extent than high-$z$ galaxies ($1.0 < z < 1.5$). This phenomenon seems to be dominated by an increase in the physical size of the signal, rather than by observational effects due to physical sizes corresponding to larger angular sizes at lower redshifts. This increase in the physical size of the signal is likely due to lower redshift galaxies having had a longer time for their hot gas, whether heated due to gravitational collapse energy or AGN feedback energy, to expand and cool.

Finally, there are several notes to make regarding the simple thermal energy models used in tSZ-SPT and tSZ-ACT (i.e. Equations \ref{eq:simEgrav} and \ref{eq:simEAGN}), based on the much more detailed Y-AGN and N-AGN simulations. At low redshifts ($z \approx 0.52$), both the Y-AGN and N-AGN $\tilde{Y}$ measurements appear to be flat with respect to mass at low masses ($M \lesssim 10^{12} M_\odot$), especially N-AGN. This is not predicted by the simple models. At high redshift ($z = 1.49$) the N-AGN $\tilde{Y}$ curve follows the model very well, while the Y-AGN curve is significantly higher than the model predicts. This is likely due to the differences in how the simple model and Horizon-AGN account for the amount of AGN feedback energy, either in magnitude or in duration. Additionally, both the Y-AGN and N-AGN $\tilde{Y}$ curves appear to systematically drop relative to the model predictions as the redshift decreases. This is likely due to the models not taking into account the radiative cooling of the galaxies and surrounding gas, which would lower the tSZ signal, and which would increase in magnitude at lower redshifts as gas has longer to cool. The simple models, as expected, do not capture the full energy dynamics over time.

Although we have tried to be as thorough and robust in this work as possible, there is a lot of further work to be done in the near future on these measurements. It may be interesting to investigate redshifts lower than 0.5 and higher than 1.5, as well as different cuts on galaxy parameters, to see if improvements can be made in these tSZ stacks, and to understand the overall evolution of the hot gas around galaxies. In addition to the stellar mass of galaxies, it could be useful to explore relations between dark matter halo masses and the tSZ effect. This work could also be expanded by doing simulations similar to Horizon-AGN but using various models of AGN feedback, in order to understand a much more detailed picture of the AGN feedback process that helps explain the observational results. With important large-scale cosmological simulations with AGN feedback continuing to be produced, it is now time for more comparisons and analyses similar to what is done in this work.

There are several next-generation CMB detectors currently running, and planned, that are potentially releasing data soon. This includes ACTPol \citep{Niemack2010}, Advanced ACTPol \citep{Calabrese2014}, SPT-Pol \citep{Austermann2012}, SPT-3G \citep{Benson2014}, and LiteBIRD \citep{Matsumura2014}, which have a large increase in sensitivity as well as more frequency bands compared to the ACT and SPT data used in the tSZ-ACT and tSZ-SPT work. These will give much better tSZ measurements, while the extra frequency bands would improve the ability to deal with contamination in the tSZ signal. 
The TolTEC camera (http://toltec.astro.umass.edu/) on the Large Millimeter Telescope, NIKA2 \citep{Calvo2016} on the IRAM 30m telescope, and MUSTANG 2 \citep{Dicker2014} on the Green Bank Telescope will measure the microwave sky at much higher angular resolution, helping to distinguish between models not only by the overall magnitude of the tSZ effect, but also  by the overall profile on the sky. These measurements can then provide future simulation work with even more robust observations to compare with, leading to increasingly constrained results.

We would like to generously thank Clotilde Laigle for making the beautiful Horizon-AGN mock lightcone image. AS would like to thank the University of Oxford Sub-department of Astrophysics for hosting him where a significant part of this work was carried out. He would also like to thank the ASU Graduate and Professional Student Association for funding part of this visit through their travel grant program. This work was carried in part within the framework of the Spin(e) grants ANR-13-BS05-0005 (http://cosmicorigin.org) of the French Agence Nationale de la Recherche and with data from the Horizon simulations (www.horizon-simulation.org). We would also like to thank the COSMOS2015 team for allowing us to use their data ahead of publication. This work is part of the Horizon-UK project. AS and ES were supported by the National Science Foundation under grant AST14-07835.


\bibliographystyle{apj}
\small
\bibliography{references}
\end{document}